\documentclass[amsmath,amssymb,aps]{revtex4}
\usepackage{natbib}
\usepackage{graphicx, epsfig}
\usepackage{latexsym}
\usepackage{epsfig}
\usepackage{latexsym}
\usepackage{amstext}
\usepackage{ulem}
\usepackage{amssymb}
\usepackage{array}
\usepackage{color}
\usepackage{graphicx}
\usepackage{dcolumn}
\usepackage{amsmath}
\usepackage{amssymb}

\newcommand{\beq}{\begin{eqnarray}}
\newcommand{\eeq}{\end{eqnarray}}
\newcommand{\beqq}{\begin{eqnarray*}}
\newcommand{\eeqq}{\end{eqnarray*}}

\newcommand{\eps}{\varepsilon}

\begin{document}
\title{Arrival time for the fastest among $N$  switching stochastic particles}
\author{S. Toste$^1$, D. Holcman$^1$}
\affiliation{$^{1}$ Group of Data Modeling, Computational Biology, IBENS, Ecole Normale Superieure-PSL, France}
\begin{abstract}
We study the first arrival time for the fastest among $N$ Brownian particles that can switch between two states inside a finite interval. The switching is modeled as a two-state Markov chain and particles can only escape in state 1.  We estimate the fastest arrival time by solving asymptotically the Fokker-Planck equations for three different initial distributions: uniformly distributed, delta-Dirac and long-tail decay. The derived formulas reveal that the fastest particle avoid switching when the switching rates are much smaller than the diffusion time scale.  The present results are compared to stochastic simulations revealing the range of validity of the derived formulas. Finally, we discuss some applications in cell biology.
\end{abstract}
\maketitle
\section{Introduction}
Key chemical reactions occurring in cells, depend on the arrival of the first molecules to small targets \cite{alberts2015essential}. This is case for a large class of agonist molecules arriving to a gated channel located on the cell membrane \cite{hille1978ionic}.  Upon arrival of the first ones, channels open and thus it is not necessarily to study the dynamics of the rest of the agonist population in the process of activation. This example shows that the statistics of the fastest arrival time is a key event in revealing the time scale of channel activation.  \\
The process of escape for the fastest is often studied as an extreme statistical event \cite{bray2013persistence,bouzigues2010mechanism,schehr2014exact}, where Brownian particles have to find a narrow window, which represents a small fraction of the explored space.  Interestingly, the probability distribution function and the mean time for the fastest, $\bar{\tau}^N$, can be computed explicitly \cite{schehr2014exact,majumdar2020extreme,basnayake2018asymptotic,basnayake2018extreme,basnayake2019fastest,Holcmanschuss2018,grebenkov2020single}. Interestingly, the asymptotic formula for the mean of the fastest depends on the initial distribution \cite{toste2021asymptotics}.

\begin{figure}[http!]
\begin{center}
\includegraphics[scale = 0.65]{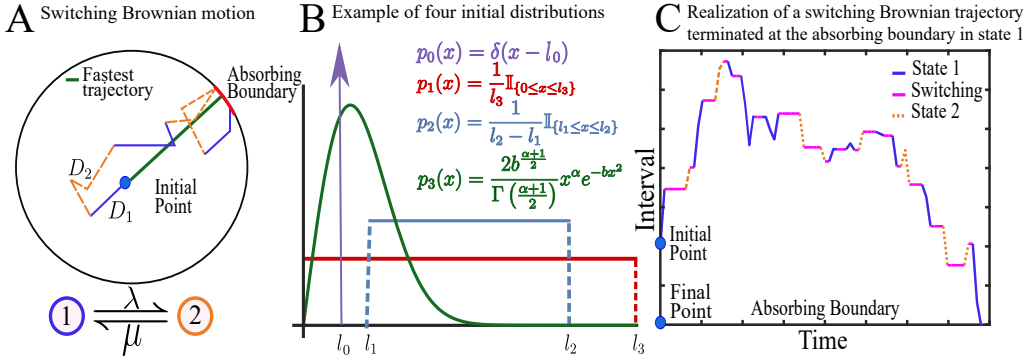}
\caption{\textbf{Schematic figure for switching Brownian dynamics.} {\textbf{A.} Example of two switching Brownian motions with two states inside a disk with a narrow absorbing boundary. The particle can escape only in state 1. The fastest trajectory (green) should move in the shortest path and be in state 1 at the absorbing boundary for escape. \textbf{B.} Four initial distributions
\textbf{C.} Realization of Brownian trajectories starting at $x=2$ and absorbed at $x=0$. }}
\label{graph1}
\end{center}
\end{figure}

We study here the extreme statistic properties for an ensemble of Brownian particles that can switch between two states characterized by two diffusion coefficient $D_1$ and $D_2$, as shown by Fig. \ref{graph1}A. The particles can only escape in state 1. Our goal is to compute the distribution and the mean arrival times for the fastest (MFAT) depending on the initial distribution.  We shall consider four different initial distributions (Fig. \ref{graph1}B). Fig. \ref{graph1}C shows a realization of a switching Brownian particle escaping  in state 1 only is. \\
The manuscript is organized as follows: Section 2 presents the stochastic model. In section 3, we develop the method to compute asymptotic solutions for two initial conditions and different diffusion coefficients. In section 4, we study the case where particles are uniformly distributed in an interval. In section 5, we present the distribution of arrival times for a long-tail initial distribution.  In the final section, we discuss applications of the present results to elementary signaling in cell biology.
\section{Stochastic model}
In the present model, we consider $N$ identical independently distributed Brownian particles that can switch at Poissonian random time  between two states as described by
\begin{eqnarray}\label{eq_reac_diff}
\huge{	1 \overset{\lambda}{\underset{\mu}{\leftrightharpoons}} 2} ,
\end{eqnarray}
with rates $\lambda$ and $\mu$. The particles can escape only in state 1. The stochastic equation for the position $x(t,i)$ in state $i$ of the particle is given for $i,j = 1,2$ by
\begin{eqnarray}
	x(t+ \Delta t,i) =\left\lbrace \begin{matrix}
		x(t,i) + \sqrt{2D_i} \Delta w_i(t) & \text{w.p $1-k_{ij}\Delta t + o(\Delta t)$} \\&\\
		x(t,j) & \text{w.p $k_{ij}\Delta t + o(\Delta t)$, $i \neq j$}
	\end{matrix},\right.
\end{eqnarray}
where $w_i(t)$ $(i = 1,2)$ are independent standard Brownian motions, $\Delta w_i(t) = w_i(t+ \Delta t) - w_i(t)$, and $k_{ij}$ are the transition rates from state $i$ to $j$. The transition probability density function $p(x,i,t|y,s,j)$ of the trajectory $x(t,i)$ with the initial condition $x(s,j) = x$, is the limit as $\Delta t \rightarrow 0$  of the integral equations
\begin{eqnarray}
	p(x,i,t + \Delta t|y,j,s) &=& \frac{1-k_{ij} \Delta t}{\sqrt{2 \pi D_i \Delta t}} \int_{\Omega}p(z,i,t|y,j,s)\exp\left\lbrace-\frac{|x-z|^2}{2D_i \Delta t}\right\rbrace dz \nonumber \\ &&\\
	&& + k_{ji}\Delta t p(x,l,t|y,j,s)+o(\Delta t) \hspace{1cm}\text{for $i,j,l = 1,2,$ $i\neq j$.} \nonumber
\end{eqnarray}
In the present case, we shall use the notation  $k_{12} = \lambda$, $k_{21} = \mu$, and $p(x,1,t|y,0) = p_1(x,t)$ and $p(x,2,t|y,0) = p_2(x,t)$. In the limit $\Delta t \rightarrow 0$, the backward system of Kolmogorov equation is given by \cite{reingruber2009gated,reingruber2009}
\begin{eqnarray}\label{backward master system2}
	\frac{\partial p_1}{\partial t}(x,t) = D_1 \frac{\partial^2 p_1}{\partial x^2}(x,t) - \lambda p_1(x,t) + \mu p_2(x,t) \nonumber \\  \nonumber\\
	\frac{\partial p_2}{\partial t}(x,t) = D_2 \frac{\partial^2 p_2}{\partial x^2}(x,t) - \mu p_2(x,t) + \lambda p_1(x,t).
\end{eqnarray}
Our goal is to find an explicit solution in the domain $[0, + \infty),$ with the boundary conditions
\begin{eqnarray}\label{boundary conditions1}
	p_1(0,t) &=& 0 \nonumber \\
	\frac{\partial p_2}{\partial x}(0,t) &=& 0.
\end{eqnarray}
When the particle starts in state $1$ at point $y>0$, the initial conditions are
given by
\begin{eqnarray} \label{initial conditions1}
p_1(x,0) &=& \delta(x-y) \nonumber \\
p_2(x,0) &=& 0.
\end{eqnarray}
We impose the normalization condition
\beq
\int_0^{\infty} p_1(x,0) + p_2(x,0) dx= 1.
\eeq
When the particle starts in state $2$ at point $y>0$, the associated initial conditions are
\begin{eqnarray} \label{initial conditions2}
	p_1(x,0) &=& 0 \nonumber \\
	p_2(x,0) &=& \delta(x-y).
\end{eqnarray}
\section{Explicit solution when the Brownian particles start in state 1}
To solve system (\ref{backward master system2}-\ref{initial conditions1}), we use the Laplace's transform in the time domain
\begin{eqnarray}
\mathcal{L}(p_1(x,t)) &=& \hat{p_1}(x,q) = \int_{0}^{\infty}p_1(x,t) \cdot e^{-qt}dt \nonumber \\
 \mathcal{L}(p_2(x,t)) &=& \hat{p_2}(x,q) = \int_{0}^{\infty}p_2(x,t) \cdot e^{-qt}dt, \nonumber
\end{eqnarray}
to get the system of two ordinary differential equations
\begin{eqnarray}
	D_1 \hat{p_1}''(x,q) -(\lambda + q)\hat{p_1}(x,q) + \mu \hat{p_2}(x,q) +\delta (x-y) &=& 0 \label{ec_p1} \\ \nonumber\\
	D_2 \hat{p_2}''(x,q) -(\mu + q)\hat{p_2}(x,q) + \lambda \hat{p_1}(x,q)  &=& 0. \label{ec que relaciona p1 y p2}
\end{eqnarray}
\subsection{Particles start in state 1 and $D_2 = 0$}
From equation (\ref{ec que relaciona p1 y p2}), when $D_2 = 0$, we have the relation
\beq \label{ec para D_2 = 0}
\hat{p}_2(x,q) = \frac{\lambda}{q + \mu}\hat{p}_1(x,q).
\eeq
Replacing this relation in equation (\ref{ec_p1}), we get
\beq \label{ec para p1 con D2 = 0}
\hat{p}_1 ''(x,q) - \frac{q(q+\theta)}{D_1(q+\mu)}\hat{p}_1(x,q) = -\frac{\delta(x-y)}{D_1},
\eeq
where $\theta  = \lambda + \mu$. We find the general solution of equation (\ref{ec para p1 con D2 = 0}),
by considering the solution $G$ of homogeneous ordinary differential equation
\begin{eqnarray}
	\frac{d^2 G}{dx^2}-\frac{q(q+\theta)}{D_1(q+\mu)}G = 0. \nonumber
\end{eqnarray}
The roots of the associated polynomial are
\begin{eqnarray} \label{roots for D2=0}
	w_{1,2} = \pm \sqrt{\frac{q(q+\theta)}{D_1(q+\mu)}},
\end{eqnarray}
and the general solution is
\begin{eqnarray}\label{general solution for D2 = 0}
	G(x,q)  = A e^{w_1 \cdot |x-y|} + B e^{w_2 \cdot |x-y|},
\end{eqnarray}
with $A,B$ $\in \mathbb{R}$. We are interested in the function that satisfies the condition
\beq
\frac{\partial G_+}{\partial x}(y,q) -\frac{\partial G_-}{\partial x}(y,q)  = -\frac{1}{D_1}. \nonumber
\eeq
Thus $A$ and $B$  satisfy the relation
\beq
2(A-B)\sqrt{\frac{q(q+\theta)}{D_1(q+\mu)}} = -\frac{1}{D_1}.\nonumber
\eeq
We can write the solution in the following form:
\begin{eqnarray}
	G(x,q) = \left(B- \sqrt{\frac{q+ \mu}{4 D_1 q (q+\theta)}}\right) e^{w_1 \cdot |x-y|} + B e^{w_2 \cdot |x-y|}  . \nonumber
\end{eqnarray}
Since the solution is  bounded, we impose $A =0$, thus
\begin{eqnarray}
	\hat{p_1}(x,q) = \sqrt{\frac{q+ \mu}{4 D_1 q (q+\theta)}} \exp \left \lbrace -\sqrt{\frac{q(q+\theta)}{D_1(q+\mu)}}\right \rbrace. \nonumber
\end{eqnarray}
To satisfy the boundary conditions $\hat{p_1}(0,q) = 0$, we finally get
\begin{eqnarray}\label{general_solution_p2_state1_D2=0}
\hat{p_1}(x,q) = \sqrt{\frac{q+ \mu}{4 D_1 q (q+\theta)}} \left( \exp \left \lbrace -\sqrt{\frac{q(q+\theta)}{D_1(q+\mu)}}|x-y|\right \rbrace -\exp \left \lbrace -\sqrt{\frac{q(q+\theta)}{D_1(q+\mu)}}|x+y|\right \rbrace \right).\nonumber
\end{eqnarray}
From relation (\ref{ec para D_2 = 0})
we obtain
\begin{eqnarray}\label{general_solution_p1_state1_D2=0}
	 \hat{p_2}(x,q) =\frac{\lambda}{q+\mu}\sqrt{\frac{q+ \mu}{4 D_1 q (q+\theta)}} \left( \exp \left \lbrace -\sqrt{\frac{q(q+\theta)}{D_1(q+\mu)}}|x-y| \right \rbrace +\exp \left \lbrace -\sqrt{\frac{q(q+\theta)}{D_1(q+\mu)}}|x+y| \right \rbrace \right).\nonumber
\end{eqnarray}
The Laplace Transform of the survival probability is given by
\beq
\hat{S}(q)= \int_{\Omega} \left( \hat{p_1}(x,q) + \hat{p_2}(x,q) \right) dx = \frac{1}{q} - \frac{q+\mu}{q(q+\theta)}\exp \left \lbrace -\sqrt{\frac{q(q+\theta)}{D_1(q+\mu)}}y \right \rbrace. \nonumber
\eeq
Using a Taylor expansion for $q$ large, we have
\beq \label{time_d2=0}
\hat{S}(q) &=& \frac{1}{q} - \frac{\exp \left \lbrace -\sqrt{\frac{q}{D_1}}y \right \rbrace}{q} + \lambda  \frac{\exp \left \lbrace -\sqrt{\frac{q}{D_1}}y \right \rbrace}{q^2}+ \lambda y \frac{\exp \left \lbrace -\sqrt{\frac{q}{D_1}}y \right \rbrace}{q \cdot \sqrt{4D_1q}} \nonumber \\
&+& O\left(\frac{\exp \left \lbrace -\sqrt{\frac{q}{D_1}}y \right \rbrace}{q^ 2}\right).\nonumber
\eeq
Applying then, the Inverse Laplace Transform, we have
for $t$ small, the approximation
\beq
S(t) &\approx& 1-\mathrm{erfc}\left(\frac{y}{\sqrt{4D_1t}}\right) + \lambda t \cdot \mathrm{erfc}\left(\frac{y}{\sqrt{4D_1t}}\right) \nonumber \\
 &\approx& 1-\sqrt{4D_1t} \frac{\exp \left \lbrace -\frac{y^2}{4D_1t} \right \rbrace}{y \sqrt{\pi}}.
\eeq
This leads for large $N$, following \cite{Basnayake2018},
\begin{eqnarray} \label{mfet_d2=0}
\bar{\tau} ^N &=& \int_{0}^{\infty}\left[S(t)\right]^N dt \approx \int_{0}^{\infty} \exp\left\lbrace \log \left\lbrace 1 - e^{-\frac{y^2}{4D_1t}} \frac{\sqrt{4D_1t}}{y \sqrt{\pi}}\right\rbrace ^N \right\rbrace dt \nonumber \\
&\approx_{N}& \frac{y^2}{4D_1\cdot \log \left(\frac{N}{\sqrt{\pi}}\right)}.
\end{eqnarray}
{
The distribution for the first arrival time in the case of $t$ small is
\beq \label{at1}
&&Pr\left\lbrace \tau^1 = t\right\rbrace = -\frac{d}{dt} S^N(t) \approx  -\frac{d}{dt}\left[\exp \left \lbrace - \frac{\sqrt{4D_1t}N}{y \sqrt{\pi}}e^{-\frac{y^2}{4D_1t}} \frac{\sqrt{4D_1t}N}{y \sqrt{\pi}}\right \rbrace \right] \nonumber\\ &=&\frac{N (\sqrt{4D_1t})}{ y\sqrt{\pi}} \exp{ \left\lbrace - \frac{y^2}{4D_1t} \right\rbrace } \exp \left \lbrace - \frac{\sqrt{4D_1t}N}{y \sqrt{\pi}}e^{-\frac{y^2}{4D_1t}} \right \rbrace \left[ \frac{y^2}{4D_1t^2} +\frac{1}{2t}\right].
\eeq
This result is similar to the one when the particles do not switch \cite{Basnayake2018}.
\begin{figure}[http!]
\begin{center}
\includegraphics[scale = 0.6]{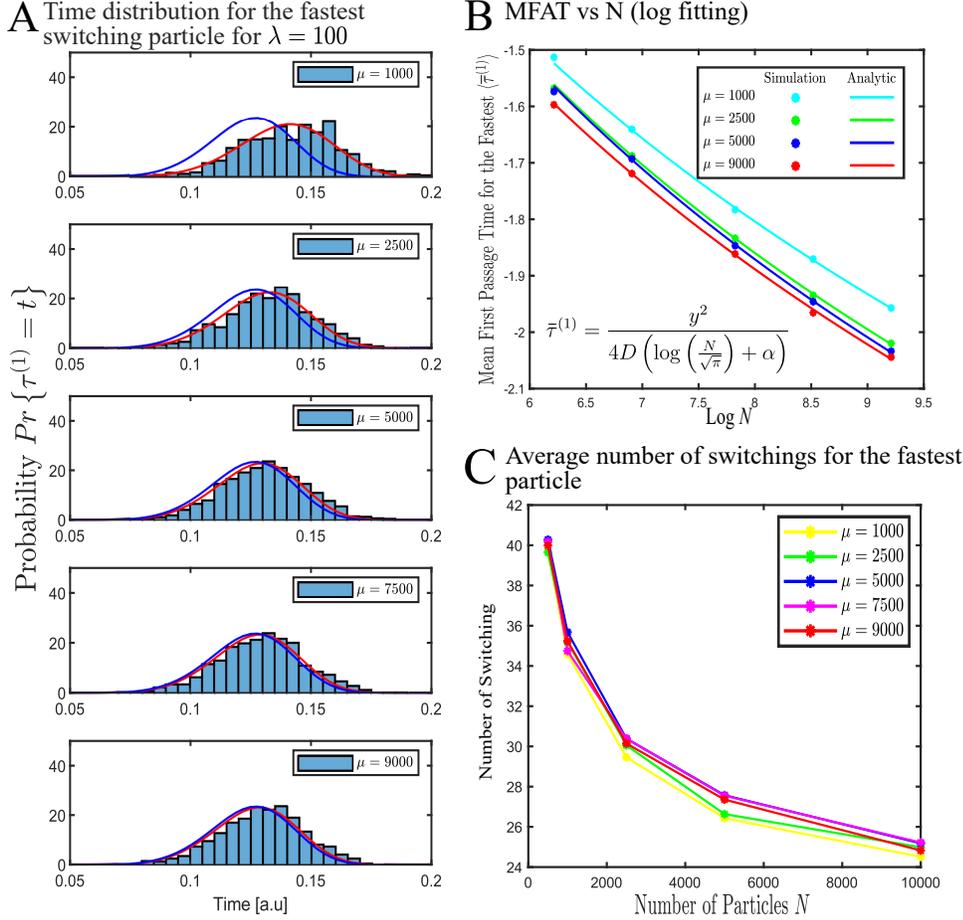}
\caption{\textbf{Mean fastest arrival time vs the number of particles $N$.} \textbf{A.} Distribution of the arrival time $\bar{\tau}^N$:  analytical short-time formula (blue) and analytical long-time formula (red) vs stochastic simulations (blue histogram) for particles starting at position $y=2$ for $N=10000$ with $1000$ runs. \textbf{B.} MFAT  vs $N$ for the stochastic simulations (colored disks) and the asymptotic formulas (continuous lines) (equation \ref{at2}) for different values of $\mu = [1000, 2500, 5000, 7500, 9000]$ and $\lambda = 100$ plotted in Log-Log scale. An optimal fit gives $\alpha = [-1.334, -0.9959, -0.9292, -0.8275, -0.8109]$. \textbf{C.} Mean number of switching  for the fastest particles.}
\label{graph3}
\end{center}
\end{figure}
When we consider large switching rates, the distribution for the first arrival time is given by the inverse of the long-time expansion of the Laplace's transform.
\beq \label{at2}
&&Pr\left\lbrace \tau^1 = t\right\rbrace = -\frac{d}{dt} S^N(t) \approx  -\frac{d}{dt}\left[\exp \left \lbrace - \frac{\sqrt{4D_1t}N\mu^{\frac{3}{2}}}{y \theta^{\frac{3}{2}} \sqrt{\pi}}e^{-\frac{y^2 \theta}{4D_1\mu t}} \frac{\sqrt{4D_1t}N}{y \sqrt{\pi}}\right \rbrace \right] \nonumber \\
&=&\frac{N \mu^{\frac{3}{2}} (\sqrt{4D_1t})}{ y\theta^{\frac{3}{2}} \sqrt{\pi}} \exp{ \left\lbrace - \frac{y^2 \theta }{4D_1 \mu t} \right\rbrace }\times \exp \left \lbrace - \frac{\sqrt{4D_1t}N\mu^{\frac{3}{2}}}{y \theta^{\frac{3}{2}} \sqrt{\pi}}e^{-\frac{y^2 \theta}{4D_1\mu t}} \right \rbrace \left[ \frac{y^2 \theta}{4D_1\mu t^2} +\frac{1}{2t}\right].
\eeq
To conclude we find that the fastest particle does not switch to the state 2 and escapes in state 1.}
\\
To evaluate the range of validity of the present formula, we decided to compare the asymptotic distributions with stochastic simulations. We generated trajectories until they reach the origin and selected the fastest one. Fig. \ref{graph3} shows the statistics for the simulation of a switching process, starting in state 1 with diffusion coefficients $D_1 = 1$ and $D_2 = 0$. We use for the initial number of particles $N = [500, 1000, 2500, 5000, 10000]$ with a time step $\Delta t = 0.0001$. The switching rates are $\lambda = 100$ and $\mu = [1000, 2500, 5000, 7500, 9000]$. We obtain a good agreement between the long-time expression for the distribution (equation \ref{at2}) of the arrival time and the stochastic simulations (Fig. \ref{graph3}A). We further confirm the decay of the MFAT vs N. We use a shift $\alpha$ in formula \ref{mfet_d2=0} to correct for possible switching: indeed the fastest particle should not switch when it starts in state 1. Indeed, we report a large number of switching (Fig. \ref{graph3}C) for the fastest particles. Interestingly, this number seems not to depend on the switching rate $\mu$. In addition,  it is unclear to us when the asymptotic limit is achieved and large should be $N$ so the number of switching read a few unit.
\subsection{Particles start in state 1 and $D_2 \neq 0$}
From  equation (\ref{ec que relaciona p1 y p2}), we obtain that
\begin{eqnarray}
	\hat{p_1}(x,q) = \frac{(\mu + q)}{\lambda}\hat{p_2}(x,q) - \frac{D_2}{\lambda}\hat{p_2}''(x,q), \nonumber
\end{eqnarray}
which can be reduced to a system of order 4
\begin{eqnarray}\label{general equation}
	\hat{p_2}^{(iv)}-\left[\frac{\lambda +q}{D_1} + \frac{\mu + q}{D_2}\right]\hat{p_2}'' + \left[\frac{(\lambda + q)(\mu + q)-\lambda \mu}{D_1 D_2}\right]\hat{p_2} = \frac{\lambda}{D_1 D_2}\delta_y.
\end{eqnarray}
The solution $G$ of the homogeneous equation is
\begin{eqnarray}
	\frac{d^4 G}{dx^4}-a\frac{d^2G}{dx^2}+bG = 0, \nonumber
\end{eqnarray}
where $a = \frac{\lambda +q}{D_1} + \frac{\mu + q}{D_2}$ and $b = \frac{(\lambda + q)(\mu + q)-\lambda \mu}{D_1 D_2}$.
The roots of the associated polynomial are
\begin{eqnarray} \label{roots}
	w_i = \pm \frac{\sqrt{a \pm \sqrt{a^2-4b}}}{\sqrt{2}} \hspace{1cm} i = \overline{1,4},
\end{eqnarray}
they are real as
\begin{eqnarray}
	a^2 - 4b = q^2\left(\frac{1}{D_1}-\frac{1}{D_2}\right)^2 + 2q\left(\frac{1}{D_1}-\frac{1}{D_2}\right)\left(\frac{\lambda}{D_1}-\frac{\mu}{D_2}\right)+\left(\frac{\lambda}{D_1}+\frac{\mu}{D_2}\right)^2, \nonumber
\end{eqnarray}
and this polynomial in $q$ is always positive as the coefficient of $q^2$ is positive and the discriminant is non positive
\begin{eqnarray}
	Disc = -16 \frac{\lambda \mu}{D_1 D_2}\left(\frac{1}{D_1}-\frac{1}{D_2}\right)^2. \nonumber
\end{eqnarray}
We have $\sqrt{a^2 - 4b} \geq 0 $. Since $b>0$ we obtain that $a - \sqrt{a^2 - 4b} > 0$, and $\sqrt{a \pm \sqrt{a^2-4b}} \in \mathbb{R}$. The general solution of equation (\ref{general equation}) is of the form
\begin{eqnarray}\label{general solution}
	G(x,q)= =A e^{w_1 \cdot |x-y|} + B e^{w_2 \cdot |x-y|} + C e^{w_3 \cdot |x-y|} + D e^{w_4 \cdot |x-y|},
\end{eqnarray}
with $A$, $B$, $C$, $D$ $\in \mathbb{R}$.
The solution is understood in the Distributions sense (\ref{general equation}).
Using the ensemble of smooth compact support $f(x)$ with $\mathbb{R}$ such that
\begin{eqnarray}
	\varphi^{(iv)} - a\varphi'' + b\varphi = f, \nonumber
\end{eqnarray}
we shall  compute $\int_{\mathbb{R}}G \cdot f dx$.
Integrating by parts and using the fact that the derivatives of $G$ until order 2 are continuous in $\mathbb{R}$, we get
\begin{eqnarray}
	\int_{\mathbb{R}}G \cdot f dx &=& \int_{\mathbb{R}}G \cdot \left(\varphi^{(iv)} - a\varphi'' + b\varphi\right) dx = \varphi(y)\left(\frac{d^3G_+}{dx^3}(y)-\frac{d^3G_-}{dx^3}(y)\right) \nonumber \\
	&=& \varphi(y)(2Aw_1^3+2Bw_2^3+2Cw_3^3+2Dw_4^3). \nonumber
\end{eqnarray}
If we choose $A$, $B$, $C$ and $D$ such that $2Aw_1^3+2Bw_2^3+2Cw_3^3+2Dw_4^3 = \frac{\lambda}{D_1 D_2}$, we got
\begin{eqnarray}
	\int_{\mathbb{R}}G \cdot f dx &=& \langle G, \varphi^{(iv)} - a\varphi'' + b\varphi \rangle = \langle  G^{(iv)} - aG'' + bG, \varphi \rangle\nonumber \\
	&=& \frac{\lambda}{D_1 D_2}\varphi(y) = \langle \frac{\lambda}{D_1 D_2} \delta (x-y), \varphi \rangle. \nonumber
\end{eqnarray}
Then,
\begin{eqnarray}
	\langle  G^{(iv)} - aG'' + bG, \varphi \rangle  &=& \langle \frac{\lambda}{D_1 D_2} \delta (x-y), \varphi \rangle \nonumber \\
	G^{(iv)} - aG'' + bG  &=&  \frac{\lambda}{D_1 D_2} \delta (x-y). \nonumber
\end{eqnarray}
Using the previous condition for $G$ be a solution in the sense of Distributions
\begin{eqnarray}
	2Aw_1^3+2Bw_2^3+2Cw_3^3+2Dw_4^3 = \frac{\lambda}{D_1 D_2}, \nonumber
\end{eqnarray}
and the condition coming from the fact that all derivatives until order 2 are continuous
\begin{eqnarray}
	2Aw_1+2Bw_2+2Cw_3+2Dw_4 = 0, \nonumber
\end{eqnarray}
we can write the solution in the following form
\begin{eqnarray}
	G(x,q) = &-&\left(\frac{Bw_2 +Cw_3 + Dw_4}{w_1}\right)\cdot e^{w_1 |x-y|}+B \cdot e^{w_2 |x-y|} \nonumber \\
	&+&\left(\frac{\lambda}{2D_1 D_2w_3(w_3^2-w_1^2)} -\frac{2Bw_2(w_2^2 - w_1^2)}{2w_3(w_3^2 - w_1^2)}-\frac{2Dw_4(w_4^2 - w_1^2)}{2w_3(w_3^2 - w_1^2)}\right)\cdot e^{w_3 |x-y|} \nonumber \\
	&+&D \cdot e^{w_4 |x-y|}. \nonumber
\end{eqnarray}
Since the solution is  bounded, we obtain that $A = C =0$, and this leads to
\begin{eqnarray}
	B = \frac{\lambda}{2D_1 D_2w_2(w_2^2-w_4^2)}, \,\,\,\, D = \frac{\lambda}{2D_1 D_2w_4(w_4^2-w_2^2)}, \nonumber
\end{eqnarray}
and the solution is
\begin{eqnarray}
	\hat{p_2}(x,q) = \frac{\lambda}{2D_1 D_2w_2(w_2^2-w_4^2)}\cdot e^{w_2 |x-y|}+ \frac{\lambda}{2D_1 D_2w_4(w_4^2-w_2^2)} \cdot e^{w_4 |x-y|}. \nonumber
\end{eqnarray}
Finally, the boundary conditions
leads to
\begin{eqnarray}\label{general_solution_p2_state1}
	\hat{p_2}(x,q) &=& \frac{\lambda}{2D_1 D_2w_2(w_2^2-w_4^2)}\cdot \left(e^{w_2 |x-y|}+e^{w_2 |x+y|}\right)\nonumber \\
	&+& \frac{\lambda}{2D_1 D_2w_4(w_4^2-w_2^2)} \cdot \left(e^{w_4 |x-y|}+e^{w_4 |x+y|}\right).
\end{eqnarray}
Using the other boundary relation (\ref{ec que relaciona p1 y p2}), we find that
\begin{eqnarray}
\hat{p_1}(x,q) = \frac{(\mu+ q - D_2w_2^2)}{2D_1 D_2w_2(w_2^2-w_4^2)}\cdot e^{w_2 |x-y|}+ \frac{(\mu + q -D_2w_4^2)}{2D_1 D_2w_4(w_4^2-w_2^2)} \cdot e^{w_4 |x-y|}, \nonumber
\end{eqnarray}
and using the boundary conditions $\hat{p_1}(0,q) = 0$, we get
\begin{eqnarray}\label{general_solution_p1_state1}
	\hat{p_1}(x,q) &=& \frac{(\mu+ q - D_2w_2^2)}{2D_1D_2w_2(w_2^2-w_4^2)}\cdot \left(e^{w_2 |x-y|} -e^{w_2 |x+y|} \right)\nonumber \\
	&+& \frac{(\mu + q -D_2w_4^2)}{2D_1 D_2w_4(w_4^2-w_2^2)} \cdot \left(e^{w_4 |x-y|} -e^{w_4 |x+y|}\right),
\end{eqnarray}
where
\begin{eqnarray} \label{root}
	w_{2,4} =-\frac{\sqrt{\frac{\lambda+q}{D_1} + \frac{\mu+q}{D_2} \pm \sqrt{q^2 \left(\frac{1}{D_1} - \frac{1}{D_2}\right)^2 + 2q \left(\frac{1}{D_1} - \frac{1}{D_2}\right)\left(\frac{\lambda}{D_1}-\frac{\mu}{D_2}\right) + \left(\frac{\lambda}{D_1} + \frac{\mu}{D_2}\right)^2}}}{\sqrt{2}}.
\end{eqnarray}
We shall now use the inverse Laplace's Transform to recover the solution for short time asymptotic.
\subsubsection{\textbf{Particles start in state 1 with the same diffusion coefficient $D_1 = D_2 = D$.}}
In the present case, the roots are
\begin{eqnarray} \label{roots in D1 = D2}
	w_2 = -\sqrt{\frac{q + \lambda +\mu}{D}}, \,\,\,\,\,\,	w_4 = -\sqrt{\frac{q}{D}},
\end{eqnarray}
and the solutions
\begin{eqnarray}
\hat{p_1}(x,q) &=& \frac{\lambda}{2 \theta \sqrt{D}\sqrt{q+\theta}}\left(e^{-\sqrt{\frac{q+\theta}{D}} |x-y|} - e^{-\sqrt{\frac{q+\theta}{D}} |x+y|}\right) \nonumber \\
&+&\frac{\mu}{2 \theta \sqrt{D}\sqrt{q}}\left(e^{-\sqrt{\frac{q}{D}} |x-y|} - e^{-\sqrt{\frac{q}{D}} |x+y|}\right), \nonumber \\
\hat{p_2}(x,q) &=& -\frac{\lambda}{2 \theta \sqrt{D}\sqrt{q+\theta}}\left(e^{-\sqrt{\frac{q+\theta}{D}} |x-y|} + e^{-\sqrt{\frac{q+\theta}{D}} |x+y|}\right) \nonumber \\
&+&\frac{\lambda}{2 \theta \sqrt{D}\sqrt{q}}\left(e^{-\sqrt{\frac{q}{D}} |x-y|} + e^{-\sqrt{\frac{q}{D}} |x+y|}\right),\nonumber
\end{eqnarray}
where $\theta  = \lambda + \mu$.
\\
The Laplace Transform of the survival probability is
\begin{eqnarray}
\hat{S}(q) &=& \int_{\Omega} \left(\hat{p_1}(x,q) + \hat{p_2}(x,q) \right)dx \nonumber\\
&=&\int_{0}^{\infty}\left(\frac{1}{\sqrt{4Dq}} e^{-\sqrt{\frac{q}{D}} |x-y|} + \frac{\lambda -\mu}{\theta \sqrt{4Dq}} e^{-\sqrt{\frac{q}{D}} |x+y|} -\frac{\lambda}{\theta \sqrt{D} \sqrt{q+\theta}} e^{-\sqrt{\frac{q+\theta}{D}} |x+y|}\right)dx, \nonumber
\end{eqnarray}
and using the integrals
\begin{itemize}
	\item $\int_{0}^{\infty} e^{-|x-y|\frac{\sqrt{q}}{\sqrt{D}}}dx = \frac{\sqrt{4D}}{\sqrt{q}}\left(2 - e^{-y \frac{\sqrt{q}}{\sqrt{D}}}\right)$
	
	\item $\int_{0}^{\infty} e^{-|x+y|\frac{\sqrt{q}}{\sqrt{D}}}dx = \frac{\sqrt{D}}{\sqrt{q}}e^{-y \frac{\sqrt{q}}{\sqrt{D}}}$,
\end{itemize}
we get,
\begin{eqnarray}
\hat{S}(q) &=& \frac{1}{q} - \frac{\mu}{\theta}\frac{e^{-y \frac{\sqrt{q}}{\sqrt{D}}}}{q}- \frac{\lambda}{\theta}\frac{e^{-y \frac{\sqrt{q+\theta}}{\sqrt{D}}}}{q+\theta}. \nonumber
\end{eqnarray}
The $q$ large expansion leads to the survival probability
\begin{eqnarray}\label{s_q_1_d1=d2}
\hat{S}(q) &=& \frac{1}{q} - \frac{e^{-y \frac{\sqrt{q}}{\sqrt{D}}}}{q} + \lambda \frac{e^{-y \frac{\sqrt{q}}{\sqrt{D}}}}{q} \left(\frac{1}{q} + \frac{y}{\sqrt{4Dq}}\right) + O\left(\frac{e^{-y \frac{\sqrt{q}}{\sqrt{D}}}}{q^{\frac{5}{2}}}\right).
\end{eqnarray}
Thus the Inverse Laplace's Transform gives
\begin{eqnarray}\label{s_t_1_d1=d2}
S(t) &\approx& \mathcal{L}^{-1}\left(\frac{1}{q}\right)- \mathcal{L}^{-1}\left(\frac{e^{-y \frac{\sqrt{q}}{\sqrt{D}}}}{q}\right) \approx 1-\mathrm{erfc}\left[\frac{y}{\sqrt{4Dt}}\right] \nonumber \\
& \approx& 1 - \frac{e^{-\frac{y^2}{4Dt}}\sqrt{4Dt}}{y \sqrt{\pi}}.
\end{eqnarray}
We obtain, thus \cite{Basnayake2018}
\begin{eqnarray}\label{tau_1_d1=d2}
	\bar{\tau} ^N &=& \int_{0}^{\infty}\left[S(t)\right]^N dt \approx \int_{0}^{\infty} \exp\left\lbrace \log \left\lbrace 1 - e^{-\frac{y^2}{4Dt}} \frac{\sqrt{4D t}}{y \sqrt{\pi}}\right\rbrace ^N \right\rbrace dt \nonumber \\
	&\approx& \frac{y^2}{4D\cdot \log \left(\frac{N}{\sqrt{\pi}}\right)}.
\end{eqnarray}
We conclude that the fastest arriving particle does not switch between  states, but escape in state 1, avoiding to change state.
\subsubsection{\textbf{Particles start in state 1 and the diffusion coefficients satisfy $D_1 \neq D_2$}}
When the diffusion coefficients differ for each state, we will use the Laplace's transform on the solutions of system (\ref{ec_p1}, \ref{ec que relaciona p1 y p2}), which will be given by (\ref{general_solution_p1_state1}) and (\ref{general_solution_p2_state1}).
To compute the Laplace transform of the survival probability, we start with
\begin{eqnarray}\label{eq de Sq para Di dif D2 empezando en D1}
\hat{S}(q) &=& \int_{0}^{\infty}(\hat{p}_1(x,q)+\hat{p}_2(x,q))dx \nonumber \\
& = & \frac{(\mu+ q - D_2w_2^2)}{2D_1D_2w_2(w_2^2-w_4^2)} \int_{0}^{\infty} \left(e^{w_2 |x-y|} -e^{w_2 |x+y|} \right) dx \nonumber \\
&+& \frac{(\mu + q -D_2w_4^2)}{2D_1 D_2w_4(w_4^2-w_2^2)} \int_{0}^{\infty} \left(e^{w_4 |x-y|} -e^{w_4 |x+y|}\right) dx \nonumber \\
&+&\frac{\lambda}{2D_1 D_2w_2(w_2^2-w_4^2)} \int_{0}^{\infty} \left(e^{w_2 |x-y|}+e^{w_2 |x+y|}\right) dx \nonumber \\
&+& \frac{\lambda}{2D_1 D_2w_4(w_4^2-w_2^2)} \int_{0}^{\infty} \left(e^{w_4 |x-y|}+e^{w_4 |x+y|}\right) dx \nonumber \\
& = & -\frac{(\mu+ q - D_2w_2^2)}{D_1D_2w_2^2(w_2^2-w_4^2)}\left(1- e^{w_2 y }\right)  -\frac{(\mu + q -D_2w_4^2)}{D_1 D_2w_4^2(w_4^2-w_2^2)}\left(1- e^{w_4 y }\right) \nonumber \\
&-&\frac{\lambda}{D_1 D_2w_2^2(w_2^2-w_4^2)}- \frac{\lambda}{D_1 D_2w_4^2(w_4^2-w_2^2)} \nonumber \\
&=& \frac{1}{q} + T_1(q) - T_2(q),
\end{eqnarray}
where $w_2$ and $w_4$ are given by formula (\ref{root}) and
\beq \label{T1 y T2}
T_1(q) = \frac{\mu + q -D_2w_2^2}{D_1 D_2w_2^2 \left(w_2^2-w_4^2\right)}e^{w_2 y}, \,\,\,\, T_2(q) = \frac{\mu + q -D_2w_4^2}{D_1 D_2w_4^2 \left(w_2^2-w_4^2\right)}e^{w_4 y}.
\eeq
Using the notations $\alpha = \frac{\lambda}{D_1} + \frac{\mu}{D_2}$, $\beta = \frac{1}{D_1}-\frac{1}{D_2}$, $\gamma = \frac{\lambda}{D_1} - \frac{\mu}{D_2}$ and $\eta = \frac{1}{D_1}+\frac{1}{D_2}$, we can rewrite $T_1(q)$ and $T_2(q)$ as
\begin{eqnarray}
T_1(q) &=& -\frac{1}{D_1 \beta q} \exp \left\lbrace-\sqrt{\frac{\eta q}{2}}y \sqrt{1+\frac{\alpha \beta}{\beta \eta q} + \frac{\beta}{\eta} \mathrm{sign}(\beta) \sqrt{1 + \frac{2\gamma}{\beta q} + \frac{\alpha ^2}{\beta^2 q^2}}} \right. \nonumber\\
&+& ln\left(1 + \frac{\gamma}{\beta q}+\mathrm{sign}(\beta) \sqrt{1 + \frac{2\gamma}{\beta q} + \frac{\alpha ^2}{\beta^2 q^2}}\right) \nonumber \\
&-& \left.ln\left(1 + \frac{2 \gamma}{\beta q}+ \frac{\alpha^2}{\beta^2 q^2}+\left(\frac{\eta}{\beta}+ \frac{\alpha}{\beta q}\right) \mathrm{sign}(\beta) \sqrt{1 + \frac{2\gamma}{\beta q} + \frac{\alpha ^2}{\beta^2 q^2}}\right) \right\rbrace, \nonumber
\end{eqnarray}
\begin{eqnarray}
T_2(q) &=& \frac{1}{D_1 \beta q} \exp \left\lbrace-\sqrt{\frac{\eta q}{2}}y \sqrt{1+\frac{\alpha \beta}{\beta \eta q} - \frac{\beta}{\eta} \mathrm{sign}(\beta) \sqrt{1 + \frac{2\gamma}{\beta q} + \frac{\alpha ^2}{\beta^2 q^2}}} \right. \nonumber\\
&+& ln\left(1 + \frac{\gamma}{\beta q}-sign(\beta) \sqrt{1 + \frac{2\gamma}{\beta q} + \frac{\alpha ^2}{\beta^2 q^2}}\right) \nonumber \\
&-& \left.ln\left(1 + \frac{2 \gamma}{\beta q}+ \frac{\alpha^2}{\beta^2 q^2}-\left(\frac{\eta}{\beta}+ \frac{\alpha}{\beta q}\right) \mathrm{sign}(\beta) \sqrt{1 + \frac{2\gamma}{\beta q} + \frac{\alpha ^2}{\beta^2 q^2}}\right) \right\rbrace. \nonumber
\end{eqnarray}
Expanding $T_1(q)$ and $T_2(q)$ for $q$ large, we have
\begin{eqnarray}
	\hat{S}(q) &=& \frac{1}{q}-\frac{1}{D_1 q}\cdot \frac{\mathrm{sign}(\beta)+1}{\beta + \eta \cdot \mathrm{sign}(\beta)} \exp \left \lbrace -\sqrt{\frac{(\eta + \beta \mathrm{sign} (\beta)) q}{2}}y\right \rbrace \nonumber \\
	&-&\frac{1}{D_1 q}\cdot \frac{\mathrm{sign}(\beta)-1}{\eta \cdot \mathrm{sign}(\beta)-\beta} \exp \left \lbrace -\sqrt{\frac{(\eta - \beta \mathrm{sign}(\beta)) q}{2}}y\right \rbrace \nonumber \\
	&+& \frac{(1 + \mathrm{sign}(\beta)) \left(1 + \frac{1}{2} \sqrt{\frac{(\eta + \beta \mathrm{sign}(\beta)) q}{2}}y \right) \left(\alpha + \gamma \cdot \mathrm{sign}(\beta)\right)}{D_1 \left(\eta + \beta \mathrm{sign}(\beta) \right) \left(\beta + \eta \cdot\mathrm{sign}(\beta)\right) q^2}\exp \left \lbrace -\sqrt{\frac{(\eta + \beta \mathrm{sign}(\beta)) q}{2}}y\right \rbrace \nonumber\\
	&+& \frac{(\mathrm{sign}(\beta) -1) \left(1 + \frac{1}{2} \sqrt{\frac{(\eta - \beta \mathrm{sign}(\beta)) q}{2}}y \right) \left(\alpha - \gamma \cdot \mathrm{sign}(\beta)\right)}{D_1 \left( \beta \mathrm{sign}(\beta) -\eta\right) \left(\beta - \eta \cdot \mathrm{sign}(\beta)\right) q^2} \exp \left \lbrace -\sqrt{\frac{(\eta - \beta \mathrm{sign}(\beta)) q}{2}}y\right \rbrace \nonumber\\
	&+& O\left(\frac{\exp \left \lbrace -\sqrt{\frac{(\eta + \beta \mathrm{sign}(\beta)) q}{2}}y\right \rbrace + \exp \left \lbrace -\sqrt{\frac{(\eta - \beta \mathrm{sign}(\beta)) q}{2}}y\right \rbrace }{q^2}\right), \nonumber
\end{eqnarray}
and for both cases, $\mathrm{sign}(\beta) = 1$ or $\mathrm{sign}(\beta) = -1$, we have,
\begin{eqnarray}
	\hat{S}(q) \approx \frac{1}{q} - \frac{e^{-y\sqrt{\frac{q}{D_1}}}}{q} + \lambda \frac{e^{-y \frac{\sqrt{q}}{\sqrt{D_1}}}}{q} \left(\frac{1}{q} + \frac{y}{\sqrt{4D_1q}}\right) + O\left(\frac{e^{-y \frac{\sqrt{q}}{\sqrt{D_1}}}}{q^{\frac{5}{2}}}\right).
\end{eqnarray}
This is,
\begin{eqnarray}
S(t) &\approx& \mathcal{L}^{-1}\left(\frac{1}{q}\right)- \mathcal{L}^{-1}\left(\frac{e^{-y \frac{\sqrt{q}}{\sqrt{D_1}}}}{q}\right) \approx 1-\mathrm{erfc}\left[\frac{y}{\sqrt{4D_1t}}\right] \nonumber \\
& \approx& 1 - \frac{e^{-\frac{y^2}{4D_1t}}\sqrt{4D_1t}}{y \sqrt{\pi}}.
\end{eqnarray}
To conclude, we obtain the same result as the case where the particles have the same diffusion coefficient (expansion (\ref{s_q_1_d1=d2}) for $D_1 = D_2$). Finally, we obtain for the mean time of the fastest the formula
\begin{eqnarray}\label{mfpt_1_d1_d2}
\bar{\tau} ^N &=& \int_{0}^{\infty}\left[S(t)\right]^N dt \approx \int_{0}^{\infty} \exp\left\lbrace \log \left\lbrace 1 - e^{-\frac{y^2}{4D_1t}} \frac{\sqrt{4D_1t}}{y \sqrt{\pi}}\right\rbrace ^N \right\rbrace dt \nonumber \\
&\approx& \frac{y^2}{4D_1\cdot \log \left(\frac{N}{\sqrt{\pi}}\right)}.
\end{eqnarray}
\subsection{Particles start in state 2}
We start with the system (\ref{backward master system2}) for which  the boundary conditions are given by (\ref{boundary conditions1}) and the initial conditions by (\ref{initial conditions2}):
\begin{eqnarray}
	p_1(x,o) &=& 0 \nonumber \\
	p_2(x,0) &=& \delta (x-y). \nonumber
\end{eqnarray}
The particles start in state 2 at point $y>0$. To determine the distribution of arrival time, we use Laplace's Transform in time to get
\begin{eqnarray}
	D_1 \hat{p_1}''(x,q) -(\lambda + q)\hat{p_1}(x,q) + \mu \hat{p_2}(x,q) &=& 0 \label{ec que relaciona p1 y p2 en el estado 2} \\ \nonumber \\
	D_2 \hat{p_2}''(x,q) -(\mu + q)\hat{p_2}(x,q) + \lambda \hat{p_1}(x,q)  +\delta (x-y)  &=& 0. \label{ec 2 system 2}
\end{eqnarray}
From equation (\ref{ec que relaciona p1 y p2 en el estado 2}) we have a relation between $\hat{p}_2(x,q)$ and $\hat{p}_1(x,q)$
\begin{eqnarray}
	\hat{p}_2(x,q) = \frac{q+ \lambda}{\mu}\hat{p}_1(x,q) - \frac{D_1}{\mu}\hat{p}_1^{''}(x,q), \nonumber
\end{eqnarray}
and replacing this in the system we get
\begin{eqnarray}\label{general solution state 2}
	\hat{p}_1^{(iv)} - \left[\frac{q+\lambda}{D_1} + \frac{q+ \mu}{D_2}\right]\hat{p}_1^{''} + \left[\frac{(q+\lambda)( q + \mu ) - \lambda \mu}{D_1 D_2}\right]\hat{p}_1 = \frac{\mu }{D_1 D_2}\delta _y.
\end{eqnarray}
This equation has the same homogeneous part as equation (\ref{general equation}), thus we need to impose
\begin{eqnarray}
	2Aw_1^3+2Bw_2^3+2Cw_3^3+2Dw_4^3 = \frac{\mu}{D_1 D_2}, \nonumber
\end{eqnarray}
for have the solution in the sense of Distributions, where $w_i$ with $i = \overline{1,4}$, are the roots of equation (\ref{root}). Also, all derivatives until second order must be continuous, then
\begin{eqnarray}
	2Aw_1+2Bw_2+2Cw_3+2Dw_4 = 0. \nonumber
\end{eqnarray}
Since the solution is bounded, we obtain that $A = C = 0$, and this leads to a solution in the form
\begin{eqnarray}
	\hat{p_1}(x,q) = \frac{\mu}{2D_1 D_2w_2(w_2^2-w_4^2)}\cdot e^{w_2 |x-y|}+ \frac{\mu}{2D_1 D_2w_4(w_4^2-w_2^2)} \cdot e^{w_4 |x-y|}. \nonumber
\end{eqnarray}
Finally, the boundary conditions imposes that
\begin{eqnarray}\label{general solution p1 state2}
	\hat{p_1}(x,q) &=& \frac{\mu}{2D_1 D_2w_2(w_2^2-w_4^2)}\cdot \left(e^{w_2 |x-y|}-e^{w_2 |x+y|}\right) \nonumber\\
	&+& \frac{\mu}{2D_1 D_2w_4(w_4^2-w_2^2)} \cdot \left(e^{w_4 |x-y|}-e^{w_4 |x+y|}\right).
\end{eqnarray}
Then, from relation (\ref{ec que relaciona p1 y p2 en el estado 2}) we obtain
\begin{eqnarray}\label{general solution p2 state 2}
	\hat{p_2}(x,q) &=& \frac{(\lambda+ q - D_1w_2^2)}{2D_1D_2w_2(w_2^2-w_4^2)}\cdot \left(e^{w_2 |x-y|} +e^{w_2 |x+y|} \right) \nonumber \\
	&+& \frac{(\lambda + q -D_1w_4^2)}{2D_1 D_2w_4(w_4^2-w_2^2)} \cdot \left(e^{w_4 |x-y|} +e^{w_4 |x+y|}\right).
\end{eqnarray}
We shall apply then, the inverse Laplace's Transform to recover the solution for short-time asymptotics.
\subsection{Solution when the particle start in state 2 for $D_1 = D_2 = D$.}

In this case we have the same roots in the form (\ref{roots in D1 = D2}) and the solutions
\begin{eqnarray}
	\hat{p_1}(x,q) &=& \frac{\mu}{\theta \sqrt{4 D q}}\left(e^{-\sqrt{\frac{q}{D}}|x-y|}-e^{-\sqrt{\frac{q}{D}}|x+y|}\right) \nonumber \\
	&-& \frac{ \mu}{\theta \sqrt{4D}\sqrt{q+ \theta}}\left(e^{-\sqrt{\frac{q +\theta}{D}}|x-y|}-e^{-\sqrt{\frac{q+ \theta}{D}}|x+y|} \right), \nonumber \\
	\hat{p_2}(x,q)&=& \frac{\mu}{\theta \sqrt{4D}\sqrt{q+\theta}}\left(e^{-\sqrt{\frac{q+\theta}{D}}|x-y|}+e^{-\sqrt{\frac{q + \theta}{D}}|x+y|}\right) \nonumber \\
	&+&\frac{\lambda}{\theta \sqrt{4Dq}}\left(e^{-\sqrt{\frac{q}{D}}|x-y|}+e^{-\sqrt{\frac{q}{D}}|x+y|}\right). \nonumber
\end{eqnarray}
The Laplace transform of the survival probability is
\begin{eqnarray}
\hat{S}(q) &=& \int_{\Omega} \left(\hat{p_1}(x,q) + \hat{p_2}(x,q) \right)dx \nonumber\\
&=&\int_{0}^{\infty}\left(\frac{1}{\sqrt{4Dq}} e^{-\sqrt{\frac{q}{D}} |x-y|} + \frac{\lambda -\mu}{\theta \sqrt{4Dq}} e^{-\sqrt{\frac{q}{D}} |x+y|} +\frac{2 \mu}{ \theta \sqrt{4D} \sqrt{q+\theta}} e^{-\sqrt{\frac{q+ \theta}{D}} |x+y|}\right)dx, \nonumber
\end{eqnarray}
and this leads to
\begin{eqnarray}
\hat{S}(q) &=& \frac{1}{q} - \frac{\mu}{\lambda + \mu}\frac{e^{-y \frac{\sqrt{q}}{\sqrt{D}}}}{q}+ \frac{\mu}{\lambda + \mu}\frac{e^{-y \frac{\sqrt{q+\theta}}{\sqrt{D}}}}{q+\theta}. \nonumber
\end{eqnarray}
Making the expansion for $q$ large, we obtain
\begin{eqnarray} \label{S2inthelimit}
\hat{S}(q) &=& \frac{1}{q} - \mu \frac{e^{-y \frac{\sqrt{q}}{\sqrt{D}}}}{q}\left(\frac{1}{q} + \frac{ y}{\sqrt{4Dq}}\right) \nonumber\\
&+& \frac{\mu \theta}{8D q^3}e^{-y \frac{\sqrt{q}}{\sqrt{D}}}\left(qy^2+ D\left(8+5\sqrt{\frac{q}{D}}y\right)\right) + O\left(\frac{e^{-y \frac{\sqrt{q}}{\sqrt{D}}}}{q^3}\right).
\end{eqnarray}
Applying the inverse Laplace's Transform, we have
\begin{eqnarray} \label{s_t_state_2}
S(t) &\approx& \mathcal{L}^{-1}\left(\frac{1}{q}\right)- \mu\mathcal{L}^{-1}\left( \frac{e^{-y \frac{\sqrt{q}}{\sqrt{D}}}}{2q^2}+  y \frac{e^{-y \frac{\sqrt{q}}{\sqrt{D}}}}{\sqrt{4Dq}q}\right) \approx 1- \mu t \cdot \mathrm{erfc}\left[\frac{y}{\sqrt{4Dt}}\right] \nonumber \\
& \approx& 1 - \mu t \frac{e^{-\frac{y^2}{4Dt}}\sqrt{4Dt}}{y \sqrt{\pi}}.
\end{eqnarray}
Considering the second order term in the expansion, we have
\begin{eqnarray}
S(t) & \approx& 1 - \mu t\left(1-\frac{\left(\lambda +\mu\right)}{2}t\right) \frac{e^{-\frac{y^2}{4Dt}}\sqrt{4Dt}}{y \sqrt{\pi}}. \nonumber
\end{eqnarray}
Thus, using the asymptotic computation of \cite{Basnayake2018}, we obtain
\begin{eqnarray} \label{mfpt2_delta}
\bar{\tau} ^N &=& \int_{0}^{\infty}\left[S(t)\right]^N dt \approx \int_{0}^{\infty}exp\left\lbrace log \left\lbrace 1 - \mu t \cdot e^{-\frac{y^2}{4Dt}} \frac{\sqrt{4D t}}{y \sqrt{\pi}}\right\rbrace ^N \right\rbrace dt \nonumber \\
&\approx& \frac{y^2}{4D\cdot \log \left(\frac{N}{\sqrt{\pi}} \cdot \mu \frac{y^2}{4D}\right)}.
\end{eqnarray}
To conclude, this result shows that the fastest particle switch only once before escape.
\subsection{Particles start in state 2 with $D_1 \neq D_2$}\label{Seccion para la Dirac Delta function en estado 2 con D1 dif de D2}
When the two diffusion coefficients are different, the general solution of system (\ref{ec 2 system 2}, \ref{ec que relaciona p1 y p2 en el estado 2}) is given by equations (\ref{general solution p1 state2}) and (\ref{general solution p2 state 2}). Laplace's transform of the survival probability leads to
\begin{eqnarray}
\hat{S}(q) &=& \int_{0}^{\infty}(\hat{p}_1(x,q)+\hat{p}_2(x,q))dx \nonumber \\
& = & \frac{\mu}{2D_1D_2w_2(w_2^2-w_4^2)} \int_{0}^{\infty} \left(e^{w_2 |x-y|} -e^{w_2 |x+y|} \right) dx \nonumber \\
&+& \frac{\mu}{2D_1 D_2w_4(w_4^2-w_2^2)} \int_{0}^{\infty} \left(e^{w_4 |x-y|} -e^{w_4 |x+y|}\right) dx \nonumber \\
&+&\frac{\lambda +q -D_1w_2^2}{2D_1 D_2w_2(w_2^2-w_4^2)} \int_{0}^{\infty} \left(e^{w_2 |x-y|}+e^{w_2 |x+y|}\right) dx \nonumber \\
&+& \frac{\lambda +q -D_1w_4^2}{2D_1 D_2w_4(w_4^2-w_2^2)} \int_{0}^{\infty} \left(e^{w_4 |x-y|}+e^{w_4 |x+y|}\right) dx \nonumber \\
& = & -\frac{\mu}{D_1D_2w_2^2(w_2^2-w_4^2)}\left(1- e^{w_2 y }\right)  -\frac{\mu }{D_1 D_2w_4^2(w_4^2-w_2^2)}\left(1- e^{w_4 y }\right) \nonumber \\
&-&\frac{\lambda+q-D_1w_2^2}{D_1 D_2w_2^2(w_2^2-w_4^2)}- \frac{\lambda +q -D_1w_4^2}{D_1 D_2w_4^2(w_4^2-w_2^2)} \nonumber \\
&=& \frac{1}{q} + T_3(q) - T_4(q), \nonumber
\end{eqnarray}
where $w_2$ and $w_4$ are given by the formula (\ref{root}), and
\beq
T_3(q) = \frac{\mu}{D_1 D_2w_2^2 \left(w_2^2-w_4^2\right)}e^{w_2 y}, \,\,\,\, T_4(q) = \frac{\mu}{D_1 D_2w_4^2 \left(w_2^2-w_4^2\right)}e^{w_4 y}.
\eeq
Rewriting $\alpha = \frac{\lambda}{D_1} + \frac{\mu}{D_2}$, $\beta = \frac{1}{D_1}-\frac{1}{D_2}$, $\gamma = \frac{\lambda}{D_1} - \frac{\mu}{D_2}$ and $\eta = \frac{1}{D_1}+\frac{1}{D_2}$, and working in $T_3(q)$ and $T_4(q)$, we obtain
\begin{eqnarray}
T_3(q) &=& \frac{2 \mu}{D_1 D_2 \beta^2 q^2} \exp \left\lbrace-\sqrt{\frac{\eta q}{2}}y \sqrt{1+\frac{\alpha \beta}{\beta \eta q} + \frac{\beta}{\eta} \mathrm{sign}(\beta) \sqrt{1 + \frac{2\gamma}{\beta q} + \frac{\alpha ^2}{\beta^2 q^2}}} \right. \nonumber\\
&-& \left. \ln\left(1 + \frac{2 \gamma}{\beta q}+ \frac{\alpha^2}{\beta^2 q^2}+\left(\frac{\eta}{\beta}+ \frac{\alpha}{\beta q}\right) \mathrm{sign}(\beta) \sqrt{1 + \frac{2\gamma}{\beta q} + \frac{\alpha ^2}{\beta^2 q^2}}\right) \right\rbrace, \nonumber
\end{eqnarray}
\begin{eqnarray}
T_4(q) &=& \frac{2 \mu}{D_1 D_2 \beta^2 q^2} \exp \left\lbrace-\sqrt{\frac{\eta q}{2}}y \sqrt{1+\frac{\alpha \beta}{\beta \eta q} - \frac{\beta}{\eta} \mathrm{sign}(\beta) \sqrt{1 + \frac{2\gamma}{\beta q} + \frac{\alpha ^2}{\beta^2 q^2}}} \right. \nonumber\\
&-& \left. \ln\left(1 + \frac{2 \gamma}{\beta q}+ \frac{\alpha^2}{\beta^2 q^2}-\left(\frac{\eta}{\beta}+ \frac{\alpha}{\beta q}\right) \mathrm{sign}(\beta) \sqrt{1 + \frac{2\gamma}{\beta q} + \frac{\alpha ^2}{\beta^2 q^2}}\right) \right\rbrace. \nonumber
\end{eqnarray}
Expanding $T_3(q)$ and $T_4(q)$ for $q$ large, we obtain for the survival probability
\begin{eqnarray} \label{desarrolloD2>D1}
\hat{S}(q) &=& \frac{1}{q}-\frac{D_2 \mu}{D_2-D_1}\frac{\exp \left \lbrace -\sqrt{\frac{q}{D_2}}y\right \rbrace}{q^2} +\frac{D_1 \mu}{D_2-D_1}\frac{\exp \left \lbrace -\sqrt{\frac{q}{D_1}}y\right \rbrace}{q^2} \nonumber \\
&+& O\left(\frac{\exp \left \lbrace -\sqrt{\frac{q}{D_1}}y\right \rbrace}{q^{\frac{5}{2}}}+\frac{\exp \left \lbrace -\sqrt{\frac{q}{D_2}}y\right \rbrace}{q^{\frac{5}{2}}}\right),
\end{eqnarray}
when $\mathrm{sign}(\beta) = 1$ ($D_2 > D_1$). Expression (\ref{desarrolloD2>D1}) contains two exponentially small terms, making the inversion difficult.  In the limit  $D_1$ close to  $D_2$, we use the  expansion $D_2 = D_1(1+\varepsilon)$ and studying the limit when $\varepsilon$ goes to zero in equation (\ref{desarrolloD2>D1}). We have
\beq
\hat{S}_{\eps}(q) &\approx& \frac{1}{q}- \frac{\mu}{q^2} \left[\exp \left \lbrace -\sqrt{\frac{q}{D_1(1+\varepsilon)}}y\right \rbrace+\frac{\exp \left \lbrace -\sqrt{\frac{q}{D_1(1+\varepsilon)}}y\right \rbrace -\exp \left \lbrace -\sqrt{\frac{q}{D_1}}y\right \rbrace}{\varepsilon}\right].\nonumber
\eeq
A Taylor expansion in  $\varepsilon$ and $\varepsilon q$ leads to
\beq
\hat{S}_{\eps}(q) = \frac{1}{q}- \frac{\mu}{q^2} \exp \left \lbrace -\sqrt{\frac{q}{D_1}}y\right \rbrace \left[1 + \sqrt{\frac{q}{4D_1}}y \right]  + O(\eps) +O\left(\eps q\right). \nonumber
\eeq
When $\eps \rightarrow 0$, the survival probability $\hat{S}_{\eps}(q)$ converges to $\hat{S}_{0}(q)$ corresponding to the solution for $D_1 = D_2$, defined by equation (\ref{s_t_state_2}). However, the convergence is not uniform in $t$ in the interval $[0,\infty[$, preventing us to use this expansion to estimate the MFAT for this case. Thus to leading order, using that
\beq
\underset{D_2 \rightarrow D_1}{\lim} Pr\left\lbrace t_1>t\right\rbrace	 \approx 1 - \mu t \frac{e^{-\frac{y^2}{4D_1t}}\sqrt{4D_1t}}{y \sqrt{\pi}}, \nonumber
\eeq
we obtain the asymptotic formula for $N\gg 1$
\begin{eqnarray}\label{MFET d2>d1}
\overline{\tau}_{\eps}^N &=& \int_{0}^{\infty}\left[S(t)\right]^N dt \approx \int_{0}^{\infty} \exp\left\lbrace \log \left\lbrace 1 - \mu \frac{e^{-\frac{y^2}{4D_1t}}\sqrt{4D_1t}}{y \sqrt{\pi}}\right\rbrace ^N \right\rbrace dt \nonumber \\
&\approx& \frac{y^2}{4D_1\cdot \log \left(\frac{N}{\sqrt{\pi}} \cdot \mu \frac{y^2}{4D_1}\right)+A_{\eps}},
\end{eqnarray}
where $A_{\eps}=A_0+\eps A_1+..$, and $A_k$ are constants. To conclude, to leading order in $\eps$, the MFAT for the case when $D_1 \neq D_2$ is similar to the case $D_1 = D_2$.\\
When $\mathrm{sign}(\beta) = -1$ (when $D_1 > D_2$), we obtain a similar expansion with the Laplace's transform of the survival probability
\begin{eqnarray}\label{desarrolloD1>D2}
\hat{S}(q) &=& \frac{1}{q}-\frac{D_1 \mu}{D_1-D_2}\frac{\exp \left \lbrace -\sqrt{\frac{q}{D_1}}y\right \rbrace}{q^2} +\frac{D_2 \mu}{D_1-D_2}\frac{\exp \left \lbrace -\sqrt{\frac{q}{D_2}}y\right \rbrace}{q^2} \nonumber \\
&+&\frac{\mu^2y \sqrt{D_1}}{2(D_1-D_2)}\frac{\exp \left \lbrace -\sqrt{\frac{q}{D_1}}y\right \rbrace}{q^{\frac{5}{2}}}-\frac{\mu^2y \sqrt{D_2}}{2(D_1-D_2)}\frac{\exp \left \lbrace -\sqrt{\frac{q}{D_2}}y\right \rbrace}{q^{\frac{5}{2}}}\nonumber\\
&+& \frac{\mu D_1^2}{(D_1-D_2)}\left(\frac{\mu}{D_1D_2} + \frac{\lambda}{D_1D_2} - \frac{2\lambda}{D_1^2}\right)\frac{\exp \left \lbrace -\sqrt{\frac{q}{D_1}}y\right \rbrace}{q^{3}} \nonumber\\
&-& \frac{\mu D_2^2}{(D_1-D_2)}\left(\frac{\mu}{D_1D_2} + \frac{\lambda}{D_1D_2} - \frac{2\mu}{D_2^2}\right)\frac{\exp \left \lbrace -\sqrt{\frac{q}{D_2}}y\right \rbrace}{q^{3}} \nonumber\\
&+& O\left(\frac{\exp \left \lbrace -\sqrt{\frac{q}{D_1}}y\right \rbrace}{q^{3}}+\frac{\exp \left \lbrace -\sqrt{\frac{q}{D_2}}y\right \rbrace}{q^{3}}\right).
\end{eqnarray}
The expression (\ref{desarrolloD1>D2}) also contains two exponentially small terms.  It is however possible to recover the case when $D_1 = D_2$, as we studied previously, by making the expansion $D_1 = D_2(1+\varepsilon)$ and studying the limit when $\varepsilon$ goes to zero in equation (\ref{desarrolloD1>D2}). In that case,  we have
\beq
\hat{S}(q) &\approx& \frac{1}{q}- \frac{\mu}{q^2} \left[\exp \left \lbrace -\sqrt{\frac{q}{D_2(1+\epsilon)}}y\right \rbrace+\frac{\exp \left \lbrace -\sqrt{\frac{q}{D_2(1+\epsilon)}}y\right \rbrace -\exp \left \lbrace -\sqrt{\frac{q}{D_2}}y\right \rbrace}{\epsilon}\right]. \nonumber
\eeq
A Taylor expansion in  $\varepsilon$ and $\varepsilon q$ leads to
\beq
\hat{S}_{\eps}(q) = \frac{1}{q}- \frac{\mu}{q^2} \exp \left \lbrace -\sqrt{\frac{q}{D_2}}y\right \rbrace \left[1 + \sqrt{\frac{q}{4D_2}}y \right]  + O(\eps) +O\left(\eps q\right). \nonumber
\eeq
When $\eps \rightarrow 0$, the survival probability $\hat{S}_{\eps}(q)$ converges to $\hat{S}_{0}(q)$ (equation (\ref{s_t_state_2})) corresponding to the solution for $D_1 = D_2$. Thus, working as we did before, we have the limit
\beq
\underset{D_1 \rightarrow D_2}{\lim} Pr\left\lbrace t_1>t\right\rbrace	 \approx 1 - \mu \frac{e^{-\frac{y^2}{4D_2t}}\sqrt{4D_2t}}{y \sqrt{\pi}}. \nonumber
\eeq
Thus to leading order, we obtain the asymptotic formula when the number of particle is large $N\gg 1$ \cite{basnayake2018extreme}
\begin{eqnarray}\label{MFET d1>d2}
\overline{\tau}_{\eps}^N &=& \int_{0}^{\infty}\left[S(t)\right]^N dt \approx \int_{0}^{\infty} \exp\left\lbrace \log \left\lbrace 1 - \mu \frac{e^{-\frac{y^2}{4D_2t}}\sqrt{4D_2t}}{y \sqrt{\pi}}\right\rbrace ^N \right\rbrace dt \nonumber \\
&\approx& \frac{y^2}{4D_2\cdot \log \left(\frac{N}{\sqrt{\pi}} \cdot \mu \frac{y^2}{4D_2}\right)+A_{\eps}},
\end{eqnarray}
where we have used the  expansion $A_{\eps}=A_0+\eps A_1+..$, and $A_k$  are constants. To conclude, we obtain an asymptotic formula for the MFAT when $D_1$ is close to $D_2$, showing that it depends on the diffusion coefficient $D_1$ and the switching rate $\mu$.
\section{Particles are initially uniformly distributed in an interval}
We now consider equation (\ref{backward master system2}) in the domain $\Omega = [0, +\infty)$ with the boundary condition (\ref{boundary conditions1}), and a uniform initial distribution in the interval $[0,y_0]$. When particles start in state 1, the initial condition is
\begin{eqnarray} \label{initial conditions3}
	p_1(x,0) &=& \frac{1}{y_0}\mathbb{I}_{\left \lbrace x\in [0,y_0]\right \rbrace} \nonumber \\
	p_2(x,0) &=& 0,
\end{eqnarray}
 and
\begin{eqnarray} \label{initial conditions4}
	p_1(x,0) &=& 0 \nonumber \\
	p_2(x,0) &=& \frac{1}{y_0}\mathbb{I}_{\left \lbrace x\in [0,y_0]\right \rbrace}.
\end{eqnarray}
when they start in state 2. We impose the normalization condition
\beq
\int_0^{\infty} p_1(x,0) + p_2(x,0) dx= 1. \nonumber
\eeq
\subsection{Particles start in state 1}
This system (\ref{general equation}) is now linear  with the indicator function as a non-homogeneous term
\begin{eqnarray}\label{general equation uniformly}
\hat{p_2}^{(iv)}-\left[\frac{\lambda +q}{D_1} + \frac{\mu + q}{D_2}\right]\hat{p_2}'' + \left[\frac{(\lambda + q)(\mu + q)-\lambda \mu}{D_1 D_2}\right]\hat{p_2} = \frac{\lambda}{y_0D_1 D_2}\mathbb{I}_{\left \lbrace x\in [0,y_0]\right \rbrace}.
\end{eqnarray}
The solution of equation (\ref{general equation uniformly}) is obtained by the convolution between the non-homogeneous function and the solution obtained with the Dirac-delta function. When $D_1 = D_2 = D$, from expansion (\ref{s_q_1_d1=d2}), we get for the survival probability,
\beq
\hat{S}(q) &\approx& \frac{1}{y_0} \int_0^{y_0} \left(\frac{1}{q} - \frac{e^{-y\sqrt{\frac{q}{D}}}}{q}\right)dy \approx \frac{1}{q}-\frac{\sqrt{D}}{y_0 q^{\frac{3}{2}}}\left(1 - e^{-y_0\sqrt{\frac{q}{D}}}\right) \approx \frac{1}{q}-\frac{\sqrt{D}}{y_0 q^{\frac{3}{2}}}. \nonumber
\eeq
The inverse Laplace leads to the short-time asymptotic,
\beq
S(t) \approx 1- \frac{\sqrt{4Dt}}{y_0 \sqrt{\pi}},
\eeq
and
\begin{eqnarray}\label{mfet_unif_1}
\bar{\tau} ^N &=& \int_{0}^{\infty}\left[S(t)\right]^N dt \approx \int_{0}^{\infty} \exp\left\lbrace \log \left\lbrace 1- \frac{\sqrt{4Dt}}{y_0 \sqrt{\pi}}\right\rbrace ^N \right\rbrace dt \nonumber \\
&\approx& \frac{y_0^2 \pi}{2DN^2}.
\end{eqnarray}
When $D_1 \neq D_2$, we obtain from  (\ref{eq de Sq para Di dif D2 empezando en D1}, \ref{T1 y T2})  the Laplace's Transform of the survival probability given by the formula
\beq
\hat{S}(q) &=& \frac{1}{y_0}  \int_0^{y_0} \left(\frac{1}{q} +\frac{\left(q +\mu-D_2w_2^2\right)e^{w_2y}}{D_1 D_2 w_2^2\left(w_2^2-w_4^2\right)} -\frac{\left(q +\mu-D_2w_4^2\right)e^{w_4y}}{D_1 D_2 w_4^2\left(w_2^2-w_4^2\right)} \right)dy \nonumber \\
&=& \frac{1}{q} -\frac{\left(q +\mu-D_2w_2^2\right)}{y_0 D_1 D_2 w_2^3\left(w_2^2-w_4^2\right)} +\frac{\left(q +\mu-D_2w_4^2\right)}{D_1 D_2 w_4^3\left(w_2^2-w_4^2\right)} \nonumber \\
&+& \frac{\left(q +\mu-D_2w_2^2\right)e^{w_2y_0}}{D_1 D_2 w_2^3\left(w_2^2-w_4^2\right)} -\frac{\left(q +\mu-D_2w_4^2\right)e^{w_4y_0}}{D_1 D_2 w_4^3\left(w_2^2-w_4^2\right)}, \nonumber
\eeq
and for $q$ large, the leading order comes from the term
\beq
\hat{S}(q) &\approx& \frac{1}{q} -\frac{\left(q +\mu-D_2w_2^2\right)}{y_0 D_1 D_2 w_2^3\left(w_2^2-w_4^2\right)} +\frac{\left(q +\mu-D_2w_4^2\right)}{D_1 D_2 w_4^3\left(w_2^2-w_4^2\right)} ,\nonumber
\eeq
due to the exponentially decaying terms.
Using te notation $T_5 (q)= -\frac{\left(q +\mu-D_2w_2^2\right)}{y_0 D_1 D_2 w_2^3\left(w_2^2-w_4^2\right)}$ and $T_6(q) = \frac{\left(q +\mu-D_2w_4^2\right)}{D_1 D_2 w_4^3\left(w_2^2-w_4^2\right)}$, and the expansion for $q$ large, we have
\beq
T_5(q) &\approx& -\frac{\sqrt{2}}{y_0D_1} \frac{1+\mathrm{sign}(\beta)}{\left(\sqrt{\beta\cdot \mathrm{sign}(\beta)+\eta}\right)^3 q^{\frac{3}{2}}},  \nonumber \\
T_6(q) &\approx& \frac{\sqrt{2}}{y_0D_1} \frac{\mathrm{sign}(\beta)-1}{\left(\sqrt{\eta-\beta \cdot \mathrm{sign}(\beta)}\right)^3 q^{\frac{3}{2}}}. \nonumber
\eeq
In both cases, when $\mathrm{sign}(\beta) = 1$ or $\mathrm{sign}(\beta)=-1$, we obtain
\beq
\hat{S}(q) \approx \frac{1}{q} - \frac{\sqrt{D_1}}{y_0q^{\frac{3}{2}}},  \nonumber
\eeq
leading to an expansion for the survival probability
\beq
S(t) \approx 1 - \frac{\sqrt{4D_1t}}{y_0 \sqrt{\pi}},
\eeq
and the asymptotic formula for the MFAT
\begin{eqnarray}
\bar{\tau} ^N &=& \int_{0}^{\infty}\left[S(t)\right]^N dt \approx \int_{0}^{\infty} \exp\left\lbrace \log \left\lbrace 1- \frac{\sqrt{4D_1t}}{y_0 \sqrt{\pi}}\right\rbrace ^N \right\rbrace dt \nonumber \\
&\approx& \frac{y_0^2 \pi}{2D_1N^2}.
\end{eqnarray}
\subsection{Particles start in state 2}
When we have the condition $D_1 = D_2$, the Laplace's transform of the survival probability is given by the formula
\beq \label{S(q) para D1 = D2 empezando unif [0,y0]}
\hat{S}(q) &\approx& \frac{1}{y_0} \int_0^{y_0} \left(\frac{1}{q} -\frac{\mu e^{-y\sqrt{\frac{q}{D}}}}{q}\left( \frac{1}{q}+\frac{y}{\sqrt{4Dq}}\right) \right)dy \nonumber \\
&\approx& \frac{1}{q} - \frac{3\mu \sqrt{D}}{2y_0q^{\frac{5}{2}}}.
\eeq
Thus, we have to leading order the formula
\begin{eqnarray}\label{mfpt_unif_2}
\bar{\tau} ^N &=& \int_{0}^{\infty}\left[S(t)\right]^N dt \approx \int_{0}^{\infty} \exp\left\lbrace \log \left\lbrace 1- \frac{\mu t \sqrt{4D_1t}}{y_0 \sqrt{\pi}}\right\rbrace ^N \right\rbrace dt \nonumber \\
&\approx& \left(\frac{\sqrt{\pi}y_0}{\mu \sqrt{4D}N}\right)^{\frac{2}{3}}\Gamma\left(\frac{5}{3}\right).
\end{eqnarray}
where $\Gamma$ is the Gamma function. When $D_1 \neq D_2$, the Laplace's transform applied to the survival probability is given by the convolution
\beq
\hat{S}(q) &=& \frac{1}{y_0} \int_{0}^{y_0}\left[ \frac{1}{q} +\frac{\mu e^{w_2y}}{D_1D_2w_2^2\left(w_2^2-w_4^2\right)}-\frac{\mu e^{w_4y}}{D_1D_2w_4^2\left(w_2^2-w_4^2\right)}\right]dy \nonumber \\
&\approx& \frac{1}{q} -\frac{\mu }{y_0D_1D_2w_2^3\left(w_2^2-w_4^2\right)}+\frac{\mu}{y_0D_1D_2w_4^3\left(w_2^2-w_4^2\right)} \nonumber \\
&=& \frac{1}{q}+T_7(q)+T_8(8), \nonumber
\eeq
where $T_7(q) =-\frac{\mu }{y_0D_1D_2w_2^3\left(w_2^2-w_4^2\right)}$ and $T_8(q) =\frac{\mu}{y_0D_1D_2w_4^3\left(w_2^2-w_4^2\right)}$. Using the expansion for $q$ large in $T_7(q)$ and $T_8(q)$, we obtain
\beq
T_7(q) \approx \frac{2\sqrt{2}\mu}{y_0 D_1 D_2 |\beta| \left(\eta + |\beta|\right)^{\frac{3}{2}} q^{\frac{5}{2}}} \text{ and }
T_8(q) \approx \frac{2\sqrt{2}\mu}{y_0 D_1 D_2 |\beta| \left(\eta - |\beta|\right)^{\frac{3}{2}} q^{\frac{5}{2}}}, \nonumber
\eeq
with $\beta,$ $\eta$ defined as in the section \ref{Seccion para la Dirac Delta function en estado 2 con D1 dif de D2}.\\
When $\beta >0$, we get
\beq
\hat{S}(q) = \frac{1}{q} - \frac{\mu D_2^{\frac{3}{2}}}{y_0 (D_2-D_1)q^{\frac{5}{2}}} +\frac{\mu D_1^{\frac{3}{2}}}{y_0 (D_2-D_1)q^{\frac{5}{2}}}.
\eeq
In the limit $D_2 = D_1(1+\varepsilon)$, an expansion in $\varepsilon$ leads to
\beq
\hat{S}_{\varepsilon}(q) &\approx& \frac{1}{q} - \frac{\mu}{y_0q^{\frac{5}{2}}} \frac{D_1^{\frac{3}{2}}\left(1+\varepsilon\right)^{\frac{3}{2}} - D_1^{\frac{3}{2}}}{D_1 \varepsilon} \nonumber \\
&\approx& \frac{1}{q} - \frac{3\mu\sqrt{D_1}}{2y_0q^{\frac{5}{2}}} - \frac{\mu \sqrt{D_1}}{y_0 q^{\frac{5}{2}}}\left(\sum_{n=1}^{\infty} \binom{\frac{3}{2}}{n+1}\varepsilon^{n}\right).
\eeq
When $\eps \rightarrow 0$, the survival probability $\hat{S}_{\eps}(q)$ converges to $\hat{S}_{0}(q)$  corresponding to the solution when $D_1 = D_2$, given by equation (\ref{S(q) para D1 = D2 empezando unif [0,y0]}). Thus to leading order, using that
\beq
\underset{D_2 \rightarrow D_1}{\lim} Pr\left\lbrace t_1>t\right\rbrace	 \approx 1 - \mu t \frac{\sqrt{4D_1t}}{y_0 \sqrt{\pi}}, \nonumber
\eeq
we obtain the asymptotic formula for $N\gg 1$
\beq
\overline{\tau}_{\eps}^N &=& \int_{0}^{\infty}\left[S(t)\right]^N dt \approx \int_{0}^{\infty} \exp\left\lbrace \log \left\lbrace 1- \frac{\mu t \sqrt{4D_1t}}{y_0 \sqrt{\pi}}\right\rbrace ^N \right\rbrace dt \nonumber \\
&\approx&  \Gamma\left(\frac{5}{3}\right)\left(\frac{y_0 \sqrt{\pi}}{\mu \sqrt{4D_1}}\right)^{\frac{2}{3}}\frac{1}{N^{\frac{2}{3}}+A_{\eps}},
\eeq
where $A_{\eps}=A_0+\eps A_1+..$, and $A_k$ are constants as before. \\
When $\beta <0$, we have then,
\beq
\hat{S}(q) = \frac{1}{q} - \frac{\mu D_1^{\frac{3}{2}}}{y_0 (D_1-D_2)q^{\frac{5}{2}}} +\frac{\mu D_2^{\frac{3}{2}}}{y_0 (D_1-D_2)q^{\frac{5}{2}}}.
\eeq
Making now, $D_1 = D_2(1+\varepsilon)$ and expanding in $\varepsilon$, we have
\beq
\hat{S}_{\varepsilon}(q) &\approx& \frac{1}{q} - \frac{\mu}{y_0q^{\frac{5}{2}}} \frac{D_2^{\frac{3}{2}}\left(1+\varepsilon\right)^{\frac{3}{2}} - D_2^{\frac{3}{2}}}{D_2 \varepsilon} \nonumber \\
&\approx& \frac{1}{q} - \frac{3\mu\sqrt{D_2}}{2y_0q^{\frac{5}{2}}} - \frac{\mu \sqrt{D_2}}{y_0 q^{\frac{5}{2}}}\left(\sum_{n=1}^{\infty} \binom{\frac{3}{2}}{n+1}\varepsilon^{n}\right).
\eeq
When $\eps \rightarrow 0$, the survival probability $\hat{S}_{\eps}(q)$ converges to $\hat{S}_{0}(q)$, corresponding to the solution when $D_1 = D_2$, given by the equation (\ref{S(q) para D1 = D2 empezando unif [0,y0]}). For te same reasons as above,  to leading order, using that
\beq
\underset{D_1 \rightarrow D_2}{\lim} Pr\left\lbrace t_1>t\right\rbrace	 \approx 1 - \mu t \frac{\sqrt{4D_2t}}{y_0 \sqrt{\pi}}, \nonumber
\eeq
we obtain  the asymptotic formula for $N\gg 1$
\beq
\overline{\tau}_{\eps}^N &=& \int_{0}^{\infty}\left[S(t)\right]^N dt \approx \int_{0}^{\infty} \exp\left\lbrace \log \left\lbrace 1- \frac{\mu t \sqrt{4D_2t}}{y_0 \sqrt{\pi}}\right\rbrace ^N \right\rbrace dt \nonumber \\
&\approx&  \Gamma\left(\frac{5}{3}\right)\left(\frac{y_0 \sqrt{\pi}}{\mu \sqrt{4D_2}}\right)^{\frac{2}{3}}\frac{1}{N^{\frac{2}{3}}+A_{\eps}},
\eeq
where $A_{\eps}=A_0+\eps A_1+..$, and $A_k$ are constants. To conclude, to leading order, we obtain a formula for the  MFAT when $D_1$ is close to $D_2$ when the initial condition starts in $[0,y_0]$.
\subsection{Particles start in state 1 uniformly distributed in $[y_1>0, y_2]$}
Considering now the initial condition $p_1(x,0) = \frac{1}{y_2-y_1}\mathbb{I}_{\left \lbrace x \in[y_1, y_2]\right \rbrace}$, when the particles start in state 1, we obtain the Laplace's transform of the survival probability given by the formula
\beq
\hat{S}(q) &=& \frac{1}{q} +\frac{\left(q +\mu-D_2w_2^2\right)\left(e^{w_2y_2}-e^{w_2y_1}\right)}{(y_2-y_1) D_1 D_2 w_2^3\left(w_2^2-w_4^2\right)} -\frac{\left(q +\mu-D_2w_4^2\right)\left(e^{w_4y_2}-e^{w_4y_1}\right)}{D_1 D_2 w_4^3\left(w_2^2-w_4^2\right)} \nonumber \\
& = & \frac{1}{q} +T_9(q) + T_{10}(q), \nonumber
\eeq
where $T_9(q) = \frac{\left(q +\mu-D_2w_2^2\right)\left(e^{w_2y_2}-e^{w_2y_1}\right)}{(y_2-y_1) D_1 D_2 w_2^3\left(w_2^2-w_4^2\right)}$ and $T_{10}(q) = -\frac{\left(q +\mu-D_2w_4^2\right)\left(e^{w_4y_2}-e^{w_4y_1}\right)}{D_1 D_2 w_4^3\left(w_2^2-w_4^2\right)}$.
And using a Taylor expansion for $q$ large,
\beq
T_9(q) &\approx& \frac{\sqrt{2}(\mathrm{sign}(b)+1)}{D_1(y_2-y_1)(\eta + |\beta|)^{\frac{3}{2}}q^{\frac{3}{2}}}\left(\exp \left \lbrace -\sqrt{\frac{(\eta + |\beta|)q}{2}}y_2\right \rbrace - \exp \left \lbrace -\sqrt{\frac{(\eta + |\beta|)q}{2}}y_1 \right \rbrace \right), \nonumber \\
T_{10}(q) &\approx& \frac{\sqrt{2}(-\mathrm{sign}(\beta)+1)}{D_1(y_2-y_1)(\eta - |\beta|)^{\frac{3}{2}}q^{\frac{3}{2}}}\left(\exp \left \lbrace -\sqrt{\frac{(\eta - |\beta|)q}{2}}y_2\right \rbrace - \exp \left \lbrace -\sqrt{\frac{(\eta - |\beta|)q}{2}}y_1 \right \rbrace \right). \nonumber
\eeq
Then, for both cases, $\mathrm{sign}(\beta) = 1$ or $\mathrm{sign}(\beta) = -1$, we have
\beq
\hat{S}(q) = \frac{1}{q} - \frac{\sqrt{D_1}}{(y_2 - y_1)q^{\frac{3}{2}}}\left(\exp \left \lbrace -\sqrt{\frac{q}{D_1}}y_1\right \rbrace - \exp \left \lbrace -\sqrt{\frac{q}{D_1}}y_2 \right \rbrace \right).
\eeq
Then, if we make $y_2 = y_1(1+\varepsilon)$, we have,
\beq
\hat{S}_{\varepsilon}(q) = \frac{1}{q} - \frac{\sqrt{D_1}}{y_1 \varepsilon q^{\frac{3}{2}}} \exp \left \lbrace -\sqrt{\frac{q}{D_1}}y_1\right \rbrace \left(1 - \exp \left \lbrace -\sqrt{\frac{q}{D_1}}y_1 \varepsilon \right \rbrace \right),
\eeq
and using teh expansion in $\varepsilon$
\beq
\hat{S}_{\varepsilon}(q) = \frac{1}{q} - \frac{\exp \left \lbrace -\sqrt{\frac{q}{D_1}}y_1\right \rbrace}{q} + \frac{\exp \left \lbrace -\sqrt{\frac{q}{D_1}}y_1\right \rbrace}{q}\sum_{n=1}^{\infty}\left(-\sqrt{\frac{q}{D_1}}y_1\varepsilon\right)^{n}.
\eeq
When $\eps \rightarrow 0$, the survival probability $S_{\eps}(t)$ converges to $S_{0}(t)$ corresponding to an initial condition for the Dirac delta function at position $y_1>0$. Once more, to leading order, using that
\beq
\underset{y_2 \rightarrow y_1}{\lim} Pr\left\lbrace t_1>t\right\rbrace	 \approx 1 - \frac{\sqrt{4D_1t}}{\sqrt{\pi}}\left[\frac{e^{-\frac{y_1^2}{4D_1t}}}{y_1}\right], \nonumber
\eeq
we obtain to leading order the asymptotic formula for $N\gg 1$
\beq \label{MFET unif [y_1,y_2] no drif}
\overline{\tau}_{\eps}^N &=& \int_{0}^{\infty}\left[S(t)\right]^N dt \approx \int_{0}^{\infty} \exp\left\lbrace \log \left\lbrace 1 - \frac{\sqrt{4D_1t}}{\sqrt{\pi}}\left[\frac{e^{-\frac{y_1^2}{4D_1t}}}{y_1}\right]\right\rbrace ^N \right\rbrace dt \nonumber \\
&\approx&   \frac{y_1^{2}}{4D_1 \,\,\log\left(\frac{N}{\sqrt{\pi}}\right)+A_{\eps}},
\eeq
where $A_{\eps}=A_0+\eps A_1+..$, and $A_k$  are constants. Here $y_1$ is the shortest distance to the absorbing boundary.
\subsection{Particles start in state 2 uniformly distributed in $[y_1, y_2]$}
In the case where the particles follows the initial condition $p_2(x,0) = \frac{1}{y_2-y_1}\mathbb{I}_{\left \lbrace x \in[y_1, y_2]\right \rbrace}$, but starts in state 2 the Laplace's transform of the survival probability given by the formula
\beq \label{S(q) para D1 dif D2 empezando unif [y_1,y2]}
\hat{S}(q) &=& \frac{1}{(y_2-y_1)} \int_{y_1}^{y_2} \left(\frac{1}{q} +\frac{\mu e^{w_2y}}{D_1D_2w_2^2\left(w_2^2 - w_4^2\right)}-\frac{\mu e^{w_4y}}{D_1D_2w_4^2\left(w_2^2 - w_4^2\right)} \right)dy \nonumber \\
&=& \frac{1}{q} +\frac{\mu \left(e^{w_2y_2}-e^{w_2y_1}\right)}{D_1D_2w_2^3\left(w_2^2 - w_4^2\right)}-\frac{\mu \left(e^{w_4y_2}-e^{w_4y_1}\right)}{D_1D_2w_4^3\left(w_2^2 - w_4^2\right)} \nonumber \\
&=& \frac{1}{q} +T_{11}(q) + T_{12}(q),
\eeq
where $T_{11}(q) = \frac{\mu \left(e^{w_2y_2}-e^{w_2y_1}\right)}{D_1D_2w_2^3\left(w_2^2 - w_4^2\right)}$ and $T_{12}(q) = -\frac{\mu \left(e^{w_4y_2}-e^{w_4y_1}\right)}{D_1D_2w_4^3\left(w_2^2 - w_4^2\right)}$, and making the Taylor expansion for $q$ large, for $T_{11}(q)$ and $T_{12}(q)$, we get
\beq
T_{11}(q) &\approx& -\frac{2\sqrt{2}\mu}{(y_2-y_1)D_1D_2|\beta|\left(\eta+|\beta|\right)^{\frac{3}{2}}q^{\frac{5}{2}}}\left(\exp \left \lbrace -\sqrt{\frac{(\eta + |\beta|)q}{2}}y_2\right \rbrace - \exp \left \lbrace -\sqrt{\frac{(\eta + |\beta|)q}{2}}y_1 \right \rbrace \right), \nonumber \\
T_{12}(q) &\approx& \frac{2\sqrt{2}\mu}{(y_2-y_1)D_1D_2|\beta|(\eta - |\beta|)^{\frac{3}{2}}q^{\frac{5}{2}}}\left(\exp \left \lbrace -\sqrt{\frac{(\eta - |\beta|)q}{2}}y_2\right \rbrace - \exp \left \lbrace -\sqrt{\frac{(\eta - |\beta|)q}{2}}y_1 \right \rbrace \right). \nonumber
\eeq
Then, for $\mathrm{sign}(\beta) = 1$, we have the expansion for the Laplace transform of the survival probability
\beq
\hat{S}(q) &=& \frac{1}{q} - \frac{\mu D_2^{\frac{3}{2}}}{(y_2-y_1)(D_2-D_1)q^{\frac{5}{2}}}\left(\exp \left \lbrace -\sqrt{\frac{q}{D_2}}y_1\right \rbrace - \exp \left \lbrace -\sqrt{\frac{q}{D_2}}y_2 \right \rbrace \right) \nonumber \\
&+& \frac{\mu D_1^{\frac{3}{2}}}{(y_2-y_1)(D_2-D_1)q^{\frac{5}{2}}}\left(\exp \left \lbrace -\sqrt{\frac{q}{D_1}}y_1\right \rbrace - \exp \left \lbrace -\sqrt{\frac{q}{D_1}}y_2 \right \rbrace \right),
\eeq
and making $y_2 = y_1(1+\varepsilon)$, we have
\beq
\hat{S}_{\varepsilon}(q) &=& \frac{1}{q} - \frac{\mu D_2^{\frac{3}{2}}}{y_1\varepsilon(D_2-D_1)q^{\frac{5}{2}}}\exp \left \lbrace -\sqrt{\frac{q}{D_2}}y_1\right \rbrace \left(1 - \exp \left \lbrace -\sqrt{\frac{q}{D_2}}y_1\varepsilon \right \rbrace \right) \nonumber \\
&+& \frac{\mu D_1^{\frac{3}{2}}}{y_1 \varepsilon(D_2-D_1)q^{\frac{5}{2}}} \exp \left \lbrace -\sqrt{\frac{q}{D_1}}y_1\right \rbrace \left(1 - \exp \left \lbrace -\sqrt{\frac{q}{D_1}}y_1\varepsilon \right \rbrace \right).
\eeq
Then, using the Taylor expansion of the exponential functions we have
\beq
\hat{S}_{\varepsilon}(q) &=& \frac{1}{q} - \frac{\mu D_2}{(D_2-D_1)q^{2}}\exp \left \lbrace -\sqrt{\frac{q}{D_2}}y_1\right \rbrace + \frac{\mu D_2}{(D_2-D_1)q^{2}} \exp \left \lbrace -\sqrt{\frac{q}{D_2}}y_1\right \rbrace \sum_{n=1}^{\infty}\left(-\sqrt{\frac{q}{D_2}}y_1 \varepsilon\right)^n \nonumber \\
&+& \frac{\mu D_1}{(D_2-D_1)q^{2}} \exp \left \lbrace -\sqrt{\frac{q}{D_1}}y_1\right \rbrace-\frac{\mu D_1}{(D_2-D_1)q^{2}} \exp \left \lbrace -\sqrt{\frac{q}{D_1}}y_1\right \rbrace \sum_{n=1}^{\infty}\left(-\sqrt{\frac{q}{D_1}}y_1 \varepsilon\right)^n. \nonumber
\eeq
When $\eps \rightarrow 0$, the survival probability $S_{\eps}(t)$ converges to $S_{0}(t)$ corresponding to an initial condition for the Dirac delta function at position $y_1$ for $D_1 \neq D_2$. Thus to leading order, using that
\beq
\underset{D_2 \rightarrow D_1}{\underset{y_2 \rightarrow y_1, }{\lim}} Pr\left\lbrace t_1>t\right\rbrace	 \approx 1 - \frac{\mu t \sqrt{4D_1t}}{\sqrt{\pi}}\left[\frac{e^{-\frac{y_1^2}{4D_1t}}}{y_1}\right], \nonumber
\eeq
we obtain to leading order the asymptotic formula for $N\gg 1$
\beq
\overline{\tau}_{\eps}^N &=& \int_{0}^{\infty}\left[S(t)\right]^N dt \approx \int_{0}^{\infty} \exp\left\lbrace \log \left\lbrace 1 - \frac{\mu t \sqrt{4D_1t}}{\sqrt{\pi}}\left[\frac{e^{-\frac{y_1^2}{4D_1t}}}{y_1}\right]\right\rbrace ^N \right\rbrace dt \nonumber \\
&\approx& \frac{y_1^2}{4D_1\cdot \log \left(\frac{N}{\sqrt{\pi}} \cdot \mu \frac{y_1^2}{4D_1}\right)+A_{\eps}},
\eeq
where $A_{\eps}=A_0+\eps A_1+..$, where $A_k$  are constants.
When $\mathrm{sign}(\beta) = -1$, we have the expansion for the Laplace transform of the survival probability
\beq
\hat{S}(q) &=& \frac{1}{q} - \frac{\mu D_1^{\frac{3}{2}}}{(y_2-y_1)(D_1-D_2)q^{\frac{5}{2}}}\left(\exp \left \lbrace -\sqrt{\frac{q}{D_1}}y_1\right \rbrace - \exp \left \lbrace -\sqrt{\frac{q}{D_1}}y_2 \right \rbrace \right) \nonumber \\
&+& \frac{\mu D_2^{\frac{3}{2}}}{(y_2-y_1)(D_1-D_2)q^{\frac{5}{2}}}\left(\exp \left \lbrace -\sqrt{\frac{q}{D_2}}y_1\right \rbrace - \exp \left \lbrace -\sqrt{\frac{q}{D_2}}y_2 \right \rbrace \right),
\eeq
and making $y_2 = y_1(1+\varepsilon)$, we have
\beq
\hat{S}_{\varepsilon}(q) &=& \frac{1}{q} - \frac{\mu D_1^{\frac{3}{2}}}{y_1\varepsilon(D_1-D_2)q^{\frac{5}{2}}}\exp \left \lbrace -\sqrt{\frac{q}{D_1}}y_1\right \rbrace \left(1 - \exp \left \lbrace -\sqrt{\frac{q}{D_1}}y_1\varepsilon \right \rbrace \right) \nonumber \\
&+& \frac{\mu D_2^{\frac{3}{2}}}{y_1 \varepsilon(D_1-D_2)q^{\frac{5}{2}}} \exp \left \lbrace -\sqrt{\frac{q}{D_2}}y_1\right \rbrace \left(1 - \exp \left \lbrace -\sqrt{\frac{q}{D_2}}y_1\varepsilon \right \rbrace \right).
\eeq
Then, using the Taylor expansion of the exponential functions we have
\beq
\hat{S}_{\varepsilon}(q) &=& \frac{1}{q} - \frac{\mu D_1}{(D_1-D_2)q^{2}}\exp \left \lbrace -\sqrt{\frac{q}{D_1}}y_1\right \rbrace + \frac{\mu D_1}{(D_1-D_2)q^{2}} \exp \left \lbrace -\sqrt{\frac{q}{D_1}}y_1\right \rbrace \sum_{n=1}^{\infty}\left(-\sqrt{\frac{q}{D_1}}y_1 \varepsilon\right)^n \nonumber \\
&+& \frac{\mu D_2}{(D_1-D_2)q^{2}} \exp \left \lbrace -\sqrt{\frac{q}{D_2}}y_1\right \rbrace-\frac{\mu D_2}{(D_1-D_2)q^{2}} \exp \left \lbrace -\sqrt{\frac{q}{D_2}}y_1\right \rbrace \sum_{n=1}^{\infty}\left(-\sqrt{\frac{q}{D_2}}y_1 \varepsilon\right)^n. \nonumber
\eeq
When $\eps \rightarrow 0$, the survival probability $S_{\eps}(t)$ converges to $S_{0}(t)$ corresponding to an initial condition for the Dirac delta function at position $y_1$ for $D_1 \neq D_2$. Using the same reasoning as above, to leading order, using that
\beq
\underset{D_1 \rightarrow D_2}{\underset{y_2 \rightarrow y_1, }{\lim}} Pr\left\lbrace t_1>t\right\rbrace	 \approx 1 - \frac{\mu t \sqrt{4D_2t}}{\sqrt{\pi}}\left[\frac{e^{-\frac{y_1^2}{4D_2t}}}{y_1}\right], \nonumber
\eeq
we obtain the asymptotic formula for $N\gg 1$
\beq
\overline{\tau}_{\eps}^N &=& \int_{0}^{\infty}\left[S(t)\right]^N dt \approx \int_{0}^{\infty} \exp\left\lbrace \log \left\lbrace 1 - \frac{\mu t \sqrt{4D_2t}}{\sqrt{\pi}}\left[\frac{e^{-\frac{y_1^2}{4D_2t}}}{y_1}\right]\right\rbrace ^N \right\rbrace dt \nonumber \\
&\approx& \frac{y_1^2}{4D_2\cdot \log \left(\frac{N}{\sqrt{\pi}} \cdot \mu \frac{y_1^2}{4D_2}\right)+A_{\eps}},
\eeq
where $A_{\eps}=A_0+\eps A_1+..$, where $A_k$  are constants.
To conclude, when $y_1>0$, this point correspond to the shortest distance to the absorbing boundary. To leading order, the MFAT is then the same as the one we found for a Dirac delta at $y_1>0$ and $D_1 = D_2$.
\section{Initial distribution with a long tail}

\subsection{Particles start in state 1}
We shall study the MFAT when the initial distribution of particles is given by
\beq
p_1(x,0) &=& \frac{2b^{\frac{1+\alpha}{2}}}{\Gamma\left(\frac{1+\alpha}{2}\right)}x^{\alpha}\exp \left \lbrace -bx^2\right \rbrace \nonumber \\
p_2(x,0) &=& 0.
\eeq
The survival probability for the switching case is given by the following expression
\beq
\hat{S}(q) &=& \frac{2b^{\frac{1+\alpha}{2}}}{\Gamma\left(\frac{1+\alpha}{2}\right)} \int_{0}^{\infty}\left[\frac{1}{q}+\frac{(\mu+q-D_2w_2^2)e^{w_2y}}{D_1D_2w_2^2\left(w_2^2-w_4^2\right)}-\frac{(\mu+q-D_2w_4^2)e^{w_4y}}{D_1D_2w_4^2\left(w_2^2-w_4^2\right)}\right]y^{\alpha}\exp \left \lbrace -bx^2\right \rbrace \nonumber \\
&=& \frac{1}{q} -\frac{\Gamma\left(1+\alpha\right)(\mu +q -D_2w_2^2)}{\Gamma\left(\frac{1+\alpha}{2}\right)2^{1+ \alpha}b^{\frac{1}{2}}D_1D_2w_2(w_2^2-w_4^2)}\mathrm{HypergeometricU}\left[1+\frac{\alpha}{2}, \frac{3}{2},\frac{w_2^2}{4b} \right] \nonumber \\
&+&\frac{\Gamma\left(1+\alpha\right)(\mu +q -D_2w_4^2)}{\Gamma\left(\frac{1+\alpha}{2}\right)2^{1+ \alpha}b^{\frac{1}{2}}D_1D_2w_4(w_2^2-w_4^2)}\mathrm{HypergeometricU}\left[1+\frac{\alpha}{2}, \frac{3}{2},\frac{w_4^2}{4b} \right],
\eeq
where, $\mathrm{HypergeometricU}(a,b,z)$ is the confluent hypergeometric function $U(a,b,z)$.
Using the large  $q$ expansion , we have
\beq
\hat{S}(q) \approx \frac{1}{q} - \frac{\Gamma\left(1+\alpha\right)}{D_1\Gamma\left(\frac{1+\alpha}{2}\right)}\left[\frac{2}{bq}\right]^{\frac{3+\alpha}{2}}\left[\frac{\mathrm{sign}(\beta)+1}{\left(\eta + |\beta|\right)^{\frac{3+\alpha}{2}}}-\frac{\mathrm{sign}(\beta)-1}{\left(\eta - |\beta|\right)^{\frac{3+\alpha}{2}}}\right]. \nonumber
\eeq
In both cases, when $\mathrm{sign}(\beta)=1$ or $\mathrm{sign}(\beta)=-1$, we have
\beq
\hat{S}(q) = \frac{1}{q} - \frac{2\Gamma \left(1+\alpha\right)D_1^{\frac{1+\alpha}{2}}}{\Gamma\left(\frac{1+\alpha}{2}\right) (bq)^{\frac{3+\alpha}{2}}}.
\eeq
Thus the inverse Laplace leads to
\beq
S(t) \approx 1-\frac{2\Gamma \left(1+\alpha\right)D_1^{\frac{1+\alpha}{2}}}{\Gamma\left(\frac{1+\alpha}{2}\right)\Gamma\left(\frac{3+\alpha}{2}\right) b^{\frac{3+\alpha}{2}}}t^{\frac{1+\alpha}{2}},
\eeq
Thus following the steps described in the previous section, we obtain that the mean time for the fastest is given by
\beq \label{mfet_lt_1}
\overline{\tau}^N &=& \int_{0}^{\infty}\left[S(t)\right]^N dt \approx \int_{0}^{\infty} \exp\left\lbrace \log \left\lbrace 1-\frac{2\Gamma \left(1+\alpha\right)D_1^{\frac{1+\alpha}{2}}}{\Gamma\left(\frac{1+\alpha}{2}\right)\Gamma\left(\frac{3+\alpha}{2}\right) b^{\frac{3+\alpha}{2}}}t^{\frac{1+\alpha}{2}}\right\rbrace ^N \right\rbrace dt \nonumber \\
&\approx&  \left[\frac{\Gamma\left(\frac{1+\alpha}{2}\right) \Gamma\left(\frac{3+\alpha}{2}\right)}{\Gamma\left(1+\alpha\right)N}\right]^{\frac{2}{1+\alpha}}\frac{b^{\frac{3+\alpha}{1+\alpha}}\Gamma\left(\frac{3+\alpha}{1+\alpha}\right)}{4^{\frac{1}{1+\alpha}}D_1}.
\eeq
\subsection{Particles start in state 2}
The particles are initially distributed in state 2
\beq
p_1(x,0) &=& 0\nonumber \\
p_2(x,0) &=&  \frac{2b^{\frac{1+\alpha}{2}}}{\Gamma\left(\frac{1+\alpha}{2}\right)}x^{\alpha}\exp \left \lbrace -bx^2\right \rbrace.
\eeq
The Laplace transform of the survival probability when $D_1 = D_2=D$ is given by
\beq
\hat{S}(q) &=& \frac{2b^{\frac{1+\alpha}{2}}}{\Gamma\left(\frac{1+\alpha}{2}\right)} \int_{0}^{\infty}\left[\frac{1}{q}-\frac{\mu e^{-y\sqrt{\frac{q}{D}}}}{\theta q }+\frac{\mu e^{-y\sqrt{\frac{q+\theta}{D}}}}{\theta \left(q + \theta \right)}\right]y^{\alpha}\exp \left \lbrace -bx^2\right \rbrace \nonumber \\
&=& \frac{1}{q} -\frac{2^{-(1+\alpha)}b^{-\frac{1}{2}}\mu \Gamma\left(1+\alpha\right)}{\Gamma\left(\frac{1+\alpha}{2}\right)\sqrt{D}}\left[\frac{1}{\sqrt{q}}\mathrm{HypergeometricU}\left[1+\frac{\alpha}{2}, \frac{3}{2},\frac{q}{4bD} \right]\right. \nonumber \\
&-& \left. \frac{1}{\sqrt{q+\theta}}\mathrm{HypergeometricU}\left[1+\frac{\alpha}{2}, \frac{3}{2},\frac{q+\theta}{4bD} \right]\right].
\eeq
where, $\mathrm{HypergeometricU}(a,b,z)$ is the confluent hypergeometric function $U(a,b,z)$ and $\theta = \lambda + \mu$.
In the large $q$ expansion, we have
\beq
\hat{S}(q) \approx \frac{1}{q} -\frac{(bD)^{\frac{1+\alpha}{2}}\mu (3+\alpha) \Gamma\left(1+\alpha\right)}{\Gamma\left(\frac{1+\alpha}{2}\right)q^{\frac{5+\alpha}{2}}},
\eeq
leading to
\beq
S(t) \approx 1 - \frac{(bD)^{\frac{1+\alpha}{2}}\mu (3+\alpha) \Gamma\left(1+\alpha\right)}{\Gamma\left(\frac{1+\alpha}{2}\right)\Gamma\left(\frac{5+\alpha}{2}\right)}t^{\frac{3+\alpha}{2}},
\eeq
and
\beq \label{mfet_lt_2}
\overline{\tau}^N  &=& \int_{0}^{\infty}\left[S(t)\right]^N dt \approx \int_{0}^{\infty} \exp\left\lbrace \log \left\lbrace 1 - \frac{(bD)^{\frac{1+\alpha}{2}}\mu (3+\alpha) \Gamma\left(1+\alpha\right)}{\Gamma\left(\frac{1+\alpha}{2}\right)\Gamma\left(\frac{5+\alpha}{2}\right)}t^{\frac{3+\alpha}{2}}\right\rbrace ^N \right\rbrace dt \nonumber \\
&\approx& \left[\frac{\Gamma\left(\frac{1+\alpha}{2}\right) \Gamma\left(\frac{5+\alpha}{2}\right)}{\mu(3+\alpha)\Gamma\left(1+\alpha\right)N}\right]^{\frac{2}{3+\alpha}}\frac{\Gamma\left(\frac{5+\alpha}{3+\alpha}\right)}{(bD)^{\frac{1+\alpha}{3+\alpha}}}.
\eeq
This decay for the mean first arrival time of the fastest particle  for $N$ large, shows a single switch before escapes. When $D_1 \neq D_2$, the Laplace transform of the survival probability gives
\beq
\hat{S}(q) &=& \frac{2b^{\frac{1+\alpha}{2}}}{\Gamma\left(\frac{1+\alpha}{2}\right)} \int_{0}^{\infty}\left[\frac{1}{q}+\frac{\mu e^{w_2y}}{D_1D_2w_2^2\left(w_2^2-w_4^2\right)}-\frac{\mu e^{w_4y}}{D_1D_2w_4^2\left(w_2^2-w_4^2\right)}\right]y^{\alpha}\exp \left \lbrace -bx^2\right \rbrace \nonumber \\
&=& \frac{1}{q} +\frac{2^{-(1+\alpha)}b^{-\frac{1}{2}}\mu \Gamma\left(1+\alpha\right)}{\Gamma\left(\frac{1+\alpha}{2}\right)D_1D_2\left(w_2^2-w_4^2\right)}\left[-\frac{1}{w_2}\mathrm{HypergeometricU}\left[1+\frac{\alpha}{2}, \frac{3}{2},\frac{w_2^2}{4b} \right]\right. \nonumber \\
&+& \left. \frac{1}{w_4}\mathrm{HypergeometricU}\left[1+\frac{\alpha}{2}, \frac{3}{2},\frac{w_4^2}{4b} \right]\right].
\eeq
Then, making the expansion for $q$ large, we can have the approximation for the survival probability
\beq
\hat{S}(q) \approx \frac{1}{q}-\frac{2\mu \Gamma \left(1+\alpha\right)}{2^{-\frac{3+\alpha}{2}}b^{-\frac{1+\alpha}{2}}\Gamma \left(\frac{1+\alpha}{2}\right) D_1 D_2 |\beta|q^{\frac{5+\alpha}{2}}}\left[ \frac{1}{\left( \eta -|\beta|\right)^{\frac{3+\alpha}{2}}} - \frac{1}{\left( \eta +|\beta|\right)^{\frac{3+\alpha}{2}}} \right],
\eeq
leading to
\beq
\hat{S}(q) \approx \frac{1}{q}-\frac{2\mu \Gamma \left(1+\alpha\right)}{b^{-\frac{1+\alpha}{2}}\Gamma \left(\frac{1+\alpha}{2}\right) (D_2-D_1) q^{\frac{5+\alpha}{2}}}\left[ D_2^{\frac{3+\alpha}{2}}-D_1^{\frac{3+\alpha}{2}} \right]
\eeq
when $\mathrm{sing}(\beta) = 1$. For $D_2 = D_1 \left(1+\varepsilon\right)$, we have
\beq
\hat{S}_{\varepsilon}(q) \approx \frac{1}{q}-\frac{2\mu \Gamma \left(1+\alpha\right)b^{\frac{1+\alpha}{2}}D_1^{\frac{3+\alpha}{2}}}{\Gamma \left(\frac{1+\alpha}{2}\right)D_1 \varepsilon q^{\frac{5+\alpha}{2}}}\left[ (1+\varepsilon)^{\frac{3+\alpha}{2}}-1 \right], \nonumber
\eeq
and expanding the above expression in $\varepsilon$, we have
\beq
\hat{S}_{\varepsilon}(q) \approx \frac{1}{q}-\frac{\mu (3+\alpha)\Gamma \left(1+\alpha\right)b^{\frac{1+\alpha}{2}}D_1^{\frac{1+\alpha}{2}}}{\Gamma \left(\frac{1+\alpha}{2}\right) q^{\frac{5+\alpha}{2}}} -\frac{2\mu \Gamma \left(1+\alpha\right)b^{\frac{1+\alpha}{2}}D_1^{\frac{1+\alpha}{2}}}{\Gamma \left(\frac{1+\alpha}{2}\right)q^{\frac{5+\alpha}{2}}}\sum_{n=1}^{\infty} \binom{\frac{3+\alpha}{2}}{n+1} \varepsilon^n. \nonumber
\eeq
When $\eps \rightarrow 0$, the survival probability $S_{\eps}(t)$ converges to $S_{0}(t)$ corresponding to an initial condition with a long tail for $D_1 = D_2$. Using the same reasoning as we did before, to leading order, using that
\beq
\underset{D_2 \rightarrow D_1, }{\lim} Pr\left\lbrace t_1>t\right\rbrace  \approx 1 - \frac{(bD_1)^{\frac{1+\alpha}{2}}\mu (3+\alpha) \Gamma\left(1+\alpha\right)}{\Gamma\left(\frac{1+\alpha}{2}\right)\Gamma\left(\frac{5+\alpha}{2}\right)}t^{\frac{3+\alpha}{2}}, \nonumber
\eeq
we obtain to leading order the asymptotic formula for $N\gg 1$
\beq
\overline{\tau}_{\eps}^N  &=& \int_{0}^{\infty}\left[S(t)\right]^N dt \approx \int_{0}^{\infty} \exp\left\lbrace \log \left\lbrace 1 - \frac{(bD_1)^{\frac{1+\alpha}{2}}\mu (3+\alpha) \Gamma\left(1+\alpha\right)}{\Gamma\left(\frac{1+\alpha}{2}\right)\Gamma\left(\frac{5+\alpha}{2}\right)}t^{\frac{3+\alpha}{2}}\right\rbrace ^N \right\rbrace dt \nonumber \\
&\approx& \left[\frac{\Gamma\left(\frac{1+\alpha}{2}\right) \Gamma\left(\frac{5+\alpha}{2}\right)}{\mu(3+\alpha)\Gamma\left(1+\alpha\right)}\right]^{\frac{2}{3+\alpha}}\frac{\Gamma\left(\frac{5+\alpha}{3+\alpha}\right)}{(bD_1)^{\frac{1+\alpha}{3+\alpha}}}\frac{1}{N^{\frac{2}{3+\alpha}} +A_{\eps}},
\eeq
where $A_{\eps}=A_0+\eps A_1+..$, where $A_k$  are constants. To conclude, to leading order in $\eps$, the MFAT when the initial distribution with a long tail and different diffusion coefficients starting in state 2 is similar to the case $D_1 = D_2$. When $\mathrm{sing}(\beta) = -1$ we have the expression for the survival probability
\beq
\hat{S}(q) \approx \frac{1}{q}-\frac{2\mu \Gamma \left(1+\alpha\right)}{b^{-\frac{1+\alpha}{2}}\Gamma \left(\frac{1+\alpha}{2}\right) (D_1-D_2) q^{\frac{5+\alpha}{2}}}\left[ D_1^{\frac{3+\alpha}{2}}-D_2^{\frac{3+\alpha}{2}} \right].
\eeq
Making $D_1 = D_2 \left(1+\varepsilon\right)$, we have
\beq
\hat{S}_{\varepsilon}(q) \approx \frac{1}{q}-\frac{2\mu \Gamma \left(1+\alpha\right)b^{\frac{1+\alpha}{2}}D_2^{\frac{3+\alpha}{2}}}{\Gamma \left(\frac{1+\alpha}{2}\right)D_2 \varepsilon q^{\frac{5+\alpha}{2}}}\left[ (1+\varepsilon)^{\frac{3+\alpha}{2}}-1 \right], \nonumber
\eeq
and expanding the above expression in $\varepsilon$, we have
\beq
\hat{S}_{\varepsilon}(q) \approx \frac{1}{q}-\frac{\mu (3+\alpha)\Gamma \left(1+\alpha\right)b^{\frac{1+\alpha}{2}}D_2^{\frac{1+\alpha}{2}}}{\Gamma \left(\frac{1+\alpha}{2}\right) q^{\frac{5+\alpha}{2}}} -\frac{2\mu \Gamma \left(1+\alpha\right)b^{\frac{1+\alpha}{2}}D_2^{\frac{1+\alpha}{2}}}{\Gamma \left(\frac{1+\alpha}{2}\right)q^{\frac{5+\alpha}{2}}}\sum_{n=1}^{\infty} \binom{\frac{3+\alpha}{2}}{n+1} \varepsilon^n. \nonumber
\eeq
When $\eps \rightarrow 0$, the survival probability $S_{\eps}(t)$ converges to $S_{0}(t)$ corresponding to an initial condition with a long tail for $D_1 = D_2$. Thus to leading order, using that
\beq
\underset{D_1 \rightarrow D_2, }{\lim} Pr\left\lbrace t_1>t\right\rbrace  \approx 1 - \frac{(bD_2)^{\frac{1+\alpha}{2}}\mu (3+\alpha) \Gamma\left(1+\alpha\right)}{\Gamma\left(\frac{1+\alpha}{2}\right)\Gamma\left(\frac{5+\alpha}{2}\right)}t^{\frac{3+\alpha}{2}}, \nonumber
\eeq
we obtain  for $N\gg 1$
\beq
\overline{\tau}_{\eps}^N  &=& \int_{0}^{\infty}\left[S(t)\right]^N dt \approx \int_{0}^{\infty} \exp\left\lbrace \log \left\lbrace 1 - \frac{(bD_2)^{\frac{1+\alpha}{2}}\mu (3+\alpha) \Gamma\left(1+\alpha\right)}{\Gamma\left(\frac{1+\alpha}{2}\right)\Gamma\left(\frac{5+\alpha}{2}\right)}t^{\frac{3+\alpha}{2}}\right\rbrace ^N \right\rbrace dt \nonumber \\
&\approx&  \left[\frac{\Gamma\left(\frac{1+\alpha}{2}\right) \Gamma\left(\frac{5+\alpha}{2}\right)}{\mu(3+\alpha)\Gamma\left(1+\alpha\right)}\right]^{\frac{2}{3+\alpha}}\frac{\Gamma\left(\frac{5+\alpha}{3+\alpha}\right)}{(bD_2)^{\frac{1+\alpha}{3+\alpha}}}\frac{1}{N^{\frac{2}{3+\alpha}} +A_{\eps}},
\eeq
where $A_{\eps}=A_0+\eps A_1+..$, where $A_k$  are constants. To conclude, the MFAT for an initial distribution with a long tail and different diffusion coefficients starting in state 2 is similar to the case $D_1 = D_2$.
\section{Discussion and concluding remarks}
In the present manuscript, we have obtained several asymptotic formulas (\ref{mfet_d2=0}, \ref{mfpt_1_d1_d2}, \ref{mfpt2_delta}, \ref{mfet_unif_1}, \ref{mfpt_unif_2},  \ref{mfet_lt_1}, \ref{mfet_lt_2}) for the mean time the fastest Brownian particles that can switch between two states escape from half a line.  These formulas are associated with different  initial distribution and whether the particle start all in state 1 or 2:  in the present model, particles escape only in state 1. Depending on the initial distribution, we found a decay in  $\frac{1}{\log N}$ when the initial distribution does not intersect the target location. However,  when $p_0(x)$ intersects the absorbing boundary, we obtain an algebraic decay with N $\frac{1}{N^{a}}$. \\
When the particles start in state 1, where escape is possible, the MFAT is similar to the formulas for the non-switching case. The main different occurs when particles start in state 2  where they can not escape and thus they need to switch at least once. We expect more interesting result when the diffusion coefficient in state $2$ is much faster than in $1$: in that case, the fastest particles should  switch to state 2, diffuse faster than in state 1 and then switch back to state 1 in order to escape. A formal derivation of this result remains to be found.\\
These extreme statistics formulas are relevant in the context of  fast molecular signaling in cell biology. This approach can be used to compute the time of activation by diffusing molecules crossing a region where switching is possible. The activation is a MFAT and depends on the main parameters, involving the geometrical organization of the domain, the initial distribution of the molecules and the specific dynamics of the particles (diffusion, switching and/or other stochastic dynamics).
\begin{figure}[http!]
\begin{center}
\includegraphics[scale = 0.65]{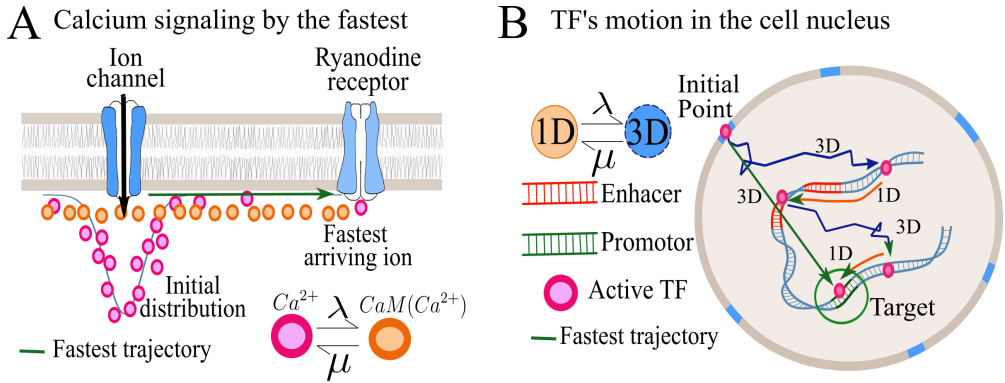}
\caption{\textbf{Application of the MFAT.} \textbf{A.} Example of Calcium ions entering through a ionic Channel. These ions can switch states between free $Ca^{2+}$ or bound to  calmodulin $CaM$, occurring diffusion. This binding interfer with a fast activation of the Ryanodine receptor \cite{basnayake2019fastest}. \textbf{B.} Example of a transcription factor $TF$ moving randomly inside a  cell nucleus, alternating between 1d and 3d diffusion. The time for the first $TF$ to activate an enhancer falls into the class of MFAT.}
\label{graph2}
\end{center}
\end{figure}
There are ubiquitous examples where particles such as molecules, transcription factors $TFs$ have to switch between different states before arriving to a small target site. For example, when particles are injected slowly in a domain,  an extended initial distribution can build up, leading to a long-tail distribution. This distribution could be approximated by a Gaussian or any other related distribution with an algebraic decay, especially when the motion can be modeled as anomalous diffusion \cite{Metzler2000}. For example, calcium ions enter in less than a few milliseconds inside a dendrite or neuronal synapses through few channels located on the membrane, as shown in Fig. \ref{graph2}A. When channels are closed again, the calcium concentration has already spread over the domain. These calcium ions can also change their state due to possible chemical reactions. One classical example is the interaction between Calcium ions and calmodulin molecules $CaM$.  Calcium signaling in dendritic spines, is often mediated by the fastest arriving calcium ion to Ryanodine receptors. During this fast process, the fastest calcium ion do not bind to a  $CaM$ molecule, as  it would lead to a much longer arrival time compared to the one observed experimentally \cite{basnayake2019fast}. This result can be explained by the present theory, showing that the fast particles do not switch.\\
Another example is the case of a transcription factor $TF$, the motion of which can be described as Brownian inside the cell nucleus. TFs can  alternate motion between a 1D sliding along the DNA and a 3D motion inside the nucleus cell (Fig. \ref{graph2}B), with diffusion coefficients $D_1\neq D_2$ and Poisonnian rates $\lambda$ and $\mu$.  Although a lot of the literature was dedicated to the case of a single TF \cite{Reingruber2011}, it is possible that the fastest TF arrives to the promoter site directly without switching, a scenario that should be further studied.
\subsection{Author contribution statement}
All authors contributed equally to the paper.

\normalem
\bibliographystyle{ieeetr}
\bibliography{BiblioCalciumSTIM1-3,references10final,ref_general,RMPbiblio4newN2,biblio,First-passage-timeBibliography,PRLCellsensingbiblio3, biblioPierre}
\end{document}